\documentclass[aps,prl,preprint,tightenlines,superscriptaddress,showpacs,byrevtex]{revtex4}
\usepackage{graphicx} 
\usepackage{dcolumn}  
\usepackage{epstopdf}
\usepackage{rotating}
\usepackage{tabularx}

\graphicspath{{ps}}

\begin{document}

\preprint{\vbox{ \hbox{BELLE Preprint 2006-26}
                             \hbox{KEK Preprint 2006-37}
                             \hbox{BELLE-CONF 0667}
}}

\title{ \quad\\[1.0cm] Moments of the electron energy spectrum and partial branching fraction of $B \to X_c e \nu$ decays at Belle}

\begin{abstract}
We report a measurement of the inclusive electron energy spectrum for charmed semileptonic decays of $B$ mesons in a  $140\,{\rm fb}^{-1}$ data sample collected at the $\Upsilon(4S)$ resonance
with the Belle detector at the KEKB asymmetric energy $e^+ e^-$ collider.  We determine the first four moments of the electron energy spectrum for threshold values of the electron energy between 0.4 and 2.0 GeV.  In addition, we provide values of the partial branching fraction (zeroth moment) for the same electron threshold energies, and independent measurements of the $B^+$ and $B^0$ partial branching fractions at 0.4 GeV and 0.6 GeV electron threshold energies.  We measure the independent $B^+$ and $B^0$ partial branching fractions with electron threshold energies of 0.4 GeV to be $\Delta \mathcal{B} (B^+ \to X_c e \nu)=(10.79 \pm 0.25({\rm stat.}) \pm 0.27({\rm sys.}))$\% and $\Delta \mathcal{B} (B^0 \to X_c e \nu)=(10.08 \pm 0.30({\rm stat.}) \pm 0.22(\rm{ sys.}))$\%.  Full correlations between all measurements are evaluated.

\end{abstract}

\affiliation{Budker Institute of Nuclear Physics, Novosibirsk}
\affiliation{Chiba University, Chiba}
\affiliation{Chonnam National University, Kwangju}
\affiliation{University of Cincinnati, Cincinnati, Ohio 45221}
\affiliation{University of Frankfurt, Frankfurt}
\affiliation{The Graduate University for Advanced Studies, Hayama, Japan} 
\affiliation{University of Hawaii, Honolulu, Hawaii 96822}
\affiliation{High Energy Accelerator Research Organization (KEK), Tsukuba}
\affiliation{Hiroshima Institute of Technology, Hiroshima}
\affiliation{University of Illinois at Urbana-Champaign, Urbana, Illinois 61801}
\affiliation{Institute of High Energy Physics, Chinese Academy of Sciences, Beijing}
\affiliation{Institute of High Energy Physics, Vienna}
\affiliation{Institute of High Energy Physics, Protvino}
\affiliation{Institute for Theoretical and Experimental Physics, Moscow}
\affiliation{J. Stefan Institute, Ljubljana}
\affiliation{Kanagawa University, Yokohama}
\affiliation{Korea University, Seoul}
\affiliation{Kyungpook National University, Taegu}
\affiliation{Swiss Federal Institute of Technology of Lausanne, EPFL, Lausanne}
\affiliation{University of Ljubljana, Ljubljana}
\affiliation{University of Maribor, Maribor}
\affiliation{University of Melbourne, Victoria}
\affiliation{Nagoya University, Nagoya}
\affiliation{Nara Women's University, Nara}
\affiliation{National Central University, Chung-li}
\affiliation{National United University, Miao Li}
\affiliation{Department of Physics, National Taiwan University, Taipei}
\affiliation{H. Niewodniczanski Institute of Nuclear Physics, Krakow}
\affiliation{Nippon Dental University, Niigata}
\affiliation{Niigata University, Niigata}
\affiliation{University of Nova Gorica, Nova Gorica}
\affiliation{Osaka City University, Osaka}
\affiliation{Osaka University, Osaka}
\affiliation{Panjab University, Chandigarh}
\affiliation{Peking University, Beijing}
\affiliation{Princeton University, Princeton, New Jersey 08544}
\affiliation{RIKEN BNL Research Center, Upton, New York 11973}
\affiliation{University of Science and Technology of China, Hefei}
\affiliation{Seoul National University, Seoul}
\affiliation{Shinshu University, Nagano}
\affiliation{Sungkyunkwan University, Suwon}
\affiliation{University of Sydney, Sydney NSW}
\affiliation{Tata Institute of Fundamental Research, Bombay}
\affiliation{Toho University, Funabashi}
\affiliation{Tohoku Gakuin University, Tagajo}
\affiliation{Tohoku University, Sendai}
\affiliation{Department of Physics, University of Tokyo, Tokyo}
\affiliation{Tokyo Institute of Technology, Tokyo}
\affiliation{Tokyo Metropolitan University, Tokyo}
\affiliation{Tokyo University of Agriculture and Technology, Tokyo}
\affiliation{Virginia Polytechnic Institute and State University, Blacksburg, Virginia 24061}
\affiliation{Yonsei University, Seoul}
  \author{P.~Urquijo}\affiliation{University of Melbourne, Victoria} 
  \author{E.~Barberio}\affiliation{University of Melbourne, Victoria} 
  \author{K.~Abe}\affiliation{High Energy Accelerator Research Organization (KEK), Tsukuba} 
  \author{K.~Abe}\affiliation{Tohoku Gakuin University, Tagajo} 
  \author{I.~Adachi}\affiliation{High Energy Accelerator Research Organization (KEK), Tsukuba} 
  \author{H.~Aihara}\affiliation{Department of Physics, University of Tokyo, Tokyo} 
  \author{D.~Anipko}\affiliation{Budker Institute of Nuclear Physics, Novosibirsk} 
  \author{V.~Aulchenko}\affiliation{Budker Institute of Nuclear Physics, Novosibirsk} 
  \author{T.~Aushev}\affiliation{Swiss Federal Institute of Technology of Lausanne, EPFL, Lausanne}\affiliation{Institute for Theoretical and Experimental Physics, Moscow} 
  \author{M.~Barbero}\affiliation{University of Hawaii, Honolulu, Hawaii 96822} 
% ...
\collaboration{The Belle Collaboration}
\noaffiliation


\maketitle

{\renewcommand{\thefootnote}{\fnsymbol{footnote}}}
\setcounter{footnote}{0}
\section{Introduction}

The Cabibbo-Kobayashi-Maskawa (CKM) matrix element $V_{cb}$ - the coupling of the $b$ quark to the $c$ quark - is a fundamental parameter of the Standard Model.  The magnitude of $V_{cb}$ can be extracted from the inclusive
decay rate of charmed semileptonic $B$-meson decays $\mathcal{B}( B \to X_c \ell \nu )$~\cite{bigi,wise}.  This paper focuses on measurements to improve the extraction of the quark mixing parameter $ |V_{cb}|$, and parameters related to the mass and kinetic energy of the $b-$quark inside the $B$ meson, $m_b$ or $\bar \Lambda$, and $\mu_\pi$ or $\lambda_1$ respectively,  from the inclusive decay spectra of charmed semileptonic $B$ meson decays. 

Several studies have shown  that the spectator model decay rate, in which bound state effects are neglected, is the leading term in a well-defined 
expansion controlled by  the parameter $\Lambda _{\rm QCD}/m_b$ \cite{gremm-kap,falk,g}. 
Non-perturbative corrections to this leading approximation arise only to order
 $1/m_b^2$. The key issue in this approach is the ability to separate non-perturbative 
 corrections (expressed as a series in powers of $1/m_b$), and perturbative 
 corrections (expressed in powers of  $\alpha _s$). 
There are various different methods 
to handle the energy scale $\mu$ used to separate long-distance from short-distance physics. 

The coefficients of the $1/m_b$ power terms are expectation values of operators that include non-perturbative physics.  In this framework, non-perturbative corrections are parameterized by quark masses and matrix elements of higher dimensional operators which are presently poorly known. 
The experimental accuracy already achieved, and that expected from larger data sets recorded by the $B$-factories, make the ensuing  theory uncertainty a major limiting factor. The extraction of the non-perturbative parameters describing the heavy 
quark masses, kinetic energy of the $b$ quark and the $1/m_b^3$ corrections directly from the data has therefore become a key issue. 

The shapes of the lepton energy spectrum and hadronic mass spectrum provide constraints on the heavy quark expansion (HQE) \cite{ref:2} based on local Operator Product Expansion (OPE) \cite{wilson}.   The non-calculable, non-perturbative quantities are parameterized in terms of expectation values of hadronic matrix elements, which can be related to the shape (characterized by moments) of inclusive decay spectra \cite{bauer-et-al-2004,ref:1}. 
Measurements of moments to high order with maximum possible phase space coverage may uncover inconsistencies in the theory. So far, measurements of the electron energy distribution have been made by the DELPHI, CLEO, BaBar and Belle collaborations \cite{ref:4,cleoel,babarel,me1}.  

The hadronic mass moments have high sensitivity to the leading order terms of the OPE.
The shape of the lepton spectrum, which is determined with greater experimental precision, is not only sensitive to leading order terms but can also constrain higher order $1/m_b$ corrections, which are the limiting factor on the precision of the theory. 
  
In this paper we report a measurement of the first four moments of the electron energy spectrum and the partial inclusive branching fractions with minimum electron energy thresholds ranging between 0.4 and 2.0 GeV in the $B$ meson rest frame.  We also provide separate measurements of  $\Delta\mathcal{B}( B^+ \to X_c e \nu )$ and $\Delta\mathcal{B}( B^0 \to X_c e \nu )~$\cite{cc} for electron energy thresholds of 0.4 GeV and 0.6 GeV.  The measurements of these independent partial branching fractions at 0.6 GeV supersede and improve upon previous results reported by the Belle Collaboration \cite{okabe}, and are the most precise measurements to date, while the measurement at 0.4 GeV sets a new lower limit of such a measurement at a $B$ factory.

 In all measurements, the choice of the lower energy endpoint is set by the limits of electron identification and prevailing backgrounds.  Only the electronic lepton channel is measured, on the basis that the precision of electron measurement is far greater than that for muons, with less material involved in the detection system.  The electron energy moments measurements are statistically limited, but not the partial branching fractions.

\section{Data sample, Detector and Simulation}

The data used in this analysis were collected with the Belle detector
at the KEKB~\cite{KEKB} asymmetric energy $e^+ e^-$ collider.
The Belle~\cite{Belle} detector is a large-solid-angle magnetic spectrometer that
consists of a three-layer silicon vertex detector (SVD), a 50-layer central drift chamber (CDC), 
an array of aerogel threshold \v{C}erenkov counters (ACC), 
a barrel-like arrangement of time-of-flight scintillation counters (TOF), and an electromagnetic calorimeter comprised of CsI(Tl) crystals (ECL) located inside a super-conducting solenoid coil that provides a 1.5~T magnetic field.  An iron flux-return located outside of the coil is instrumented to detect $K_L^0$ mesons and to identify muons (KLM).

The present results are based on a $140\,{\rm fb}^{-1}$ data sample collected at the $\Upsilon (4S)$ resonance (on-resonance), which contains $1.52 \times 10^8$ $B \overline B$ pairs.  An additional $15\,{\rm fb}^{-1}$ data sample taken at 60MeV below the $\Upsilon (4S)$ resonance (off-resonance) is used to perform subtraction of background arising from the continuum $e^+e^- \rightarrow q \bar q$ process.  Events are selected by  fully reconstructing one of the $B$ mesons, produced in pairs from $\Upsilon (4S)$ decays.

We use Monte Carlo (MC) techniques to simulate the production and decay of $B$ mesons, and the detector response.  The simulated sample of generic $B \overline B$ events is equivalent to three times the on-resonance integrated luminosity.  
In addition we use a simulated sample of $B \to X_u \ell \nu$ events equivalent to 25 times the expected rate in data.  Simulated events are generated with the EVTGEN event generator~\cite{evtgen} and processed through the Belle detector simulation based on GEANT~\cite{BelleMC}.  

For the simulation of $B \rightarrow X_c e \nu$ decays, we have chosen a variety of models.  For $B \to D e \nu$ and $B \to D^* e \nu$ decays we use parameterizations~\cite{HQET,CLN,GL}  of the form factors, based on heavy quark effective theory (HQET).  Decays to pseudoscalar mesons are described by a single form factor $F_D(w)/F_D(1)=1-\rho^2_D(w-1)$, where the variable $w$ is the scalar product of the $B$ and $D$ meson four-vector velocities.  We use the world average value of the slope parameter $\rho_D^2=1.17\pm0.18$~\cite{HFAG}.  The rate for $B \to D^* e \nu$ can be described by three amplitudes, which depend on three parameters denoted $\rho^2$, $R_1$ and $R_2$.  We adopt the world average value, $\rho^2=1.19\pm0.06$~\cite{HFAG} and the most recently measured values for $R_1=1.396\pm0.075$ and $R_2=0.885\pm0.047$ \cite{dstarbabar}.  The branching fractions of the $D$ and $D^*$ components are based on values reported in the Review of Particle Physics \cite{PDG}.

Details of the various decays to higher mass $D^{**}$ resonances are less well known.  The $D^{**} e \nu$ component includes both narrow orbitally excited charmed mesons and broad resonances.  The existence of both the broad and narrow resonant states is well established~\cite{kuzmin}, however, only the narrow state semileptonic branching fractions have been measured~\cite{LEP}, with limits placed on the broad state branching fractions.
Decay shape characteristics of these states in semileptonic $B$ decays have not been measured and must be estimated from theory predictions.  We use the model by Leibovich {\it et al.} ~\cite{LLSW} (LLSW).   Differential decay rates are predicted for various resonant $D^{**} e \nu$ decays, using limits from measurements to resonant states, (semileptonic~\cite{LEP} and hadronic~\cite{kuzmin}), as well as the full rate to $D^{(*)} \pi e \nu$ states~\cite{matsumoto}, and full inclusive rates~\cite{PDG}.
These limits enable an estimate of the $1/m_Q$ corrections to the currently used ISGW2 decay models~\cite{ISGW2}.  
The uncertainty on the measured $D^{**} e \nu$ resonances, in conjunction with the theoretical estimates provide bounds on the differential decay rates (and branching fractions) of the $D^{**} e \nu$ contributions.   We have adopted a prescription by Goity and Roberts \cite{ref:9} for the non-resonant $B \to D^{(*)} \pi e \nu$ decay shapes.
 
 The MC sample used to model background $b \to u$ events is a hybrid mix of inclusive and exclusive contributions. The exclusive channels $\pi$, $\rho$ and $\omega$ decays are produced with the SLPOLE model~\cite{evtgen}. 
 Other resonant semileptonic decays (charged $a_{0,1,2}$, $b_1$ for neutral $B$ and neutral $\eta$, $\eta'$, $a_{0,1,2}$, $b_1$, $f_{0,1,2}$ for charged $B$) are simulated with the ISGW2 model~\cite{ISGW2}.  
Contributions from the inclusive part of the mix are implemented with the shape function parameterization (defined in Ref.~\cite{dfn}).
The inclusive branching fraction is set to the world average value,  $\mathcal{B}(B \to X_u \ell \nu) = (2.16 \pm 0.33) \times 10^{-3}$ ~\cite{HFAG}.

\section{Event Selection}
We first identify hadronic events based on charged track multiplicity and total visible energy, suppressing backgrounds from QED, $e^+e^-\to \tau^+\tau^-$, and beam-gas events. The selection procedure is described in detail elsewhere ~\cite{hadron}. We then fully reconstruct one $B$ meson in one of several hadronic modes to determine its charge, flavor, and momentum, referred to as the ``tag-side"  $B$ ($B_{\rm{tag}}$).
The $B_{\rm{tag}}$ candidates are reconstructed in the decay modes $B^+ \to \bar D^{(*)0} \pi^+, \bar D^{(*)0} \rho^+, \bar D^{(*)0} a_1^+$ and $B^0 \to D^{(*)-} \pi^+, D^{(*)-} \rho^+, D^{(*)-} a_1^+$, yielding a high purity $B$ meson sample. The following sub-decay modes of the charmed meson are
reconstructed:

\begin{itemize}
\item $\bar{D}^{*0}\to \bar{D}^0\pi^0, \bar{D}^0\gamma$,
\item $D^{*-}\to \bar{D}^0\pi^-, D^-\pi^0$,
\item $\bar{D}^0\to K^+\pi^-, K^+\pi^-\pi^0, K^+\pi^-\pi^-\pi^+, K_S\pi^+\pi^-, K^0_S\pi^0$ and
\item $D^-\to K^+\pi^-\pi^+, K^0_S\pi$.
\end{itemize}

For each selected event, we calculate the beam-energy constrained mass, $M_{\mathrm{bc}}$, and the energy difference, $\Delta E$:
\begin{equation}
  M_{\mathrm{bc}} = \sqrt{(E^*_{\mathrm{beam}})^2-(p^*_B)^2}, \quad \Delta E =
  E^*_B-E^*_{\mathrm{beam}},
\end{equation}
where $E^*_{\mathrm{beam}}$, $p^*_B$ and $E^*_B$ are the beam energy, the
reconstructed $B$ momentum and the reconstructed $B$ energy in the centre of mass
frame, respectively. 
Events with   $5.27~\mathrm{GeV/}c^2<M_{\mathrm{bc}}<5.29~\mathrm{GeV/}c^2$  and $-0.06~\mathrm{GeV}<\Delta E< 0.08~\mathrm{GeV}$ are considered for further analysis.
In this region the $B_{\rm tag}$  purity is 66\% (72\%) for $B^+$ ($B^0$) tags.
The number of $B^+$  and $B^0$  candidates (with statistical errors) in the  signal region ($N_{\rm{tag}}$),  after continuum and combinatorial background (due to incorrect reconstruction or tagging of the tagged $B$) subtraction, is $63185\pm621$ and $39504\pm392$, respectively.  Subtraction of these backgrounds is performed with the same method as used for prompt electron events, and is described in detail later.

\section{Electron Selection and electron momentum reconstruction}

\subsection{Identification and Selection}
We search for electrons produced by semileptonic $B$ decays on the ``non-tag'' side. 
Electron candidates are required to originate from near the interaction vertex and pass through the barrel region of the detector, corresponding to an angular acceptance of $35^{\circ}<\theta_{\rm lab}<125^{\circ}$, where $\theta_{\rm lab}$ denotes the polar angle of the electron candidate with respect to the direction opposite to the positron beam. We exclude tracks used in the reconstruction of the $B_{\rm{tag}}$ and multiple reconstructed tracks generated by   low-momentum particles spiralling in the drift chamber. 

Electron candidates are selected on the basis of the ratio of the energy detected in the ECL to the track momentum, matching between the positions of the charged track and ECL cluster, the ECL shower shape, the energy loss in the drift chamber and the response of the ACC \cite{Belle}. 
In events with multiple identified electrons, only the highest momentum electron is considered as an electron candidate.   The electron identification efficiency and the probabilities to misidentify a pion, kaon or proton as an electron have been measured as a function of the laboratory momentum and angles.  The average electron identification efficiency and hadron misidentification rate are 97\% and 0.7\% respectively, over the full phase space.

\subsection{Bremsstrahlung recovery}

Due to the emission of highly energetic photons from electrons,  the determination of electron momenta solely from reconstructed track information results in the reconstructed momenta being softer than expected.
To alleviate this, the momentum of each electron is determined using additional information from the ECL.  Neutral clusters of energy below 1 GeV contained within a cone of 0.05 radians around the electron track direction from the interaction point are added to the electron energy.  The radius of the cone around the electron has been chosen to maximize the signal to noise ratio for photons emitted by electrons.  The photon energy cut optimizes the electron energy resolution, as over-correction for photon radiation causes a significant bias of the reconstructed momentum.

Photon radiation may be due to either bremsstrahlung radiation in the detector material in front of the ECL, or to QED radiation in the decay process.  Simulation of detector bremsstrahlung in the detector material in front of the ECL requires that the description of the material be very precise.   The method of summing all radiated photons in proximity to the track decreases the dependence on the accuracy of the material description.  Prompt photons due to high order QED corrections are accounted for in the MC with the use of the PHOTOS package \cite{PHOTOS}.

The electron momenta are calculated in the $B$ meson rest frame ($p^{*B}_{e}$), exploiting the knowledge of the momentum of the fully reconstructed $B$. We require $p^{*B}_{e} \geq 0.4$~GeV/$c$. 
The stated selection criteria result in an efficiency of 45$-$65\% for selecting $B \to X_c e \nu$ decays, which is dependent on the electron energy.

\section{Background Subtraction}

The reconstructed electron momentum  spectrum is contaminated by background processes, which are  evaluated and subtracted from the distribution before the extraction of the moments.  Contamination of the spectrum is predominantly due to continuum background, combinatorial background,
 cascade charm decays $b \rightarrow c \rightarrow q \ell \nu$ (secondary),  $J/ \psi, \psi (2S)$, Dalitz decays, photon conversions,  fake electrons and  $B \to X_u \ell \nu$ decays.  These will be described in turn.

\subsection{Non-$B \overline{B}$ Background}
The shape of the continuum background is derived from off-resonance data, and is normalized using the off- to on-resonance luminosity ratio and cross section difference.   
The statistical uncertainty of the continuum normalization factor  is determined by the number of detected Bhabha events used for the measurement of the integrated luminosity.
There are very few events in the off-resonance data that pass the event and particle selection criteria so we choose to fit an exponential to the off-resonance $p^{*B}_e$ distribution before renormalizing, of the form $f(\vec{a},p^{*B}_e)=\exp{(a_1 + a_2 p^{*B}_e)} $, where $\vec{a}$ are the set of free parameters in the fit.

\subsection{$B \overline{B}$ Background}
In the charged $B$-meson sample, prompt semileptonic decays ($b \to q \ell \nu$) of the ``non-tag" side $B$ mesons are separated from cascade charm decays ($b \to c \to q \ell \nu$), based on the correlation between the flavor of the tagging $B$ and the electron charge.  In neutral $B$-meson decays, mixing may occur, flipping the correlation.  Thus in the neutral $B$ sample we do not require this correlation.

\begin{figure}[htb!]
  \begin{tabular}{cc}
\includegraphics[width=0.48\textwidth]{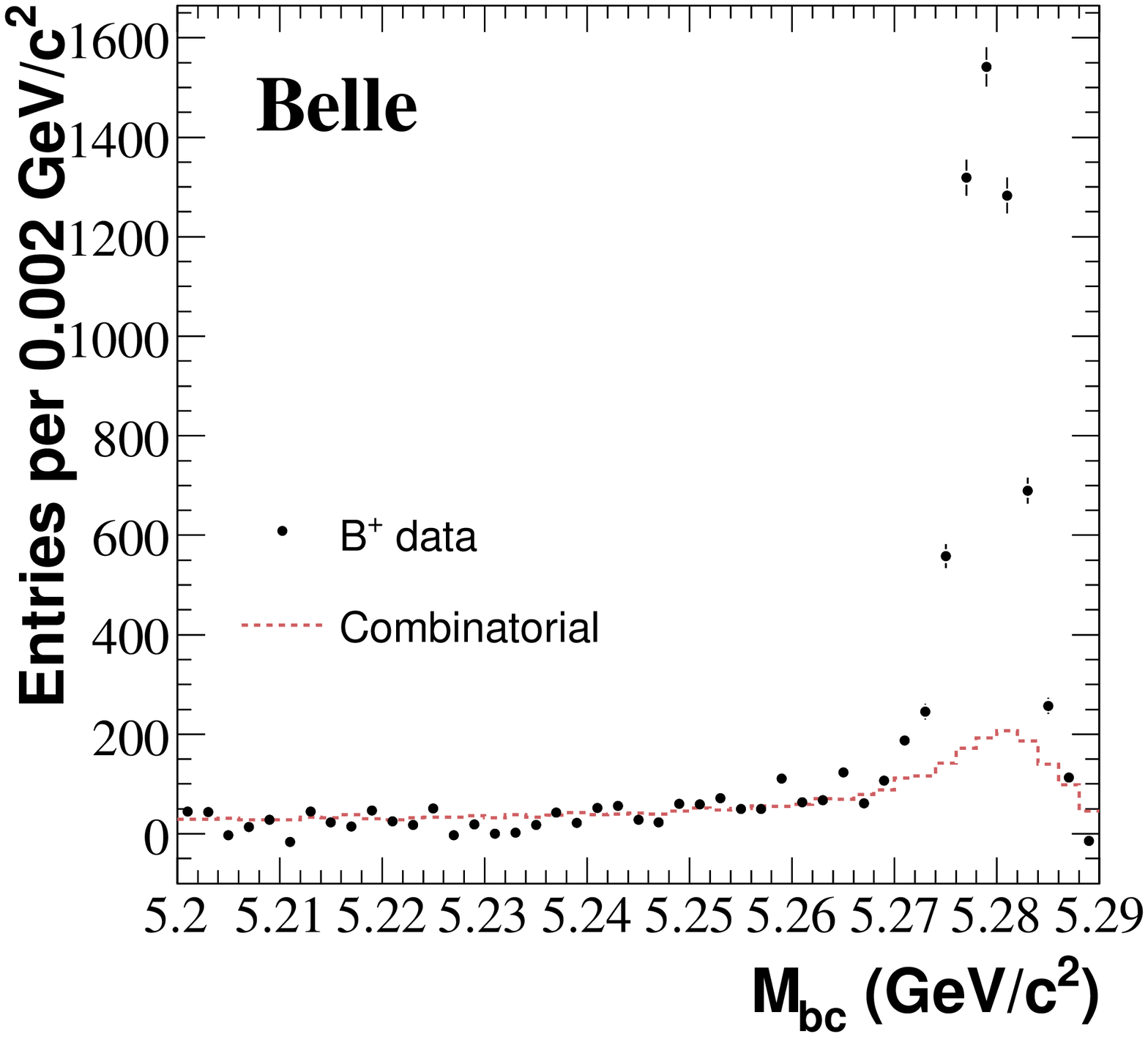}& 
\includegraphics[width=0.48\textwidth]{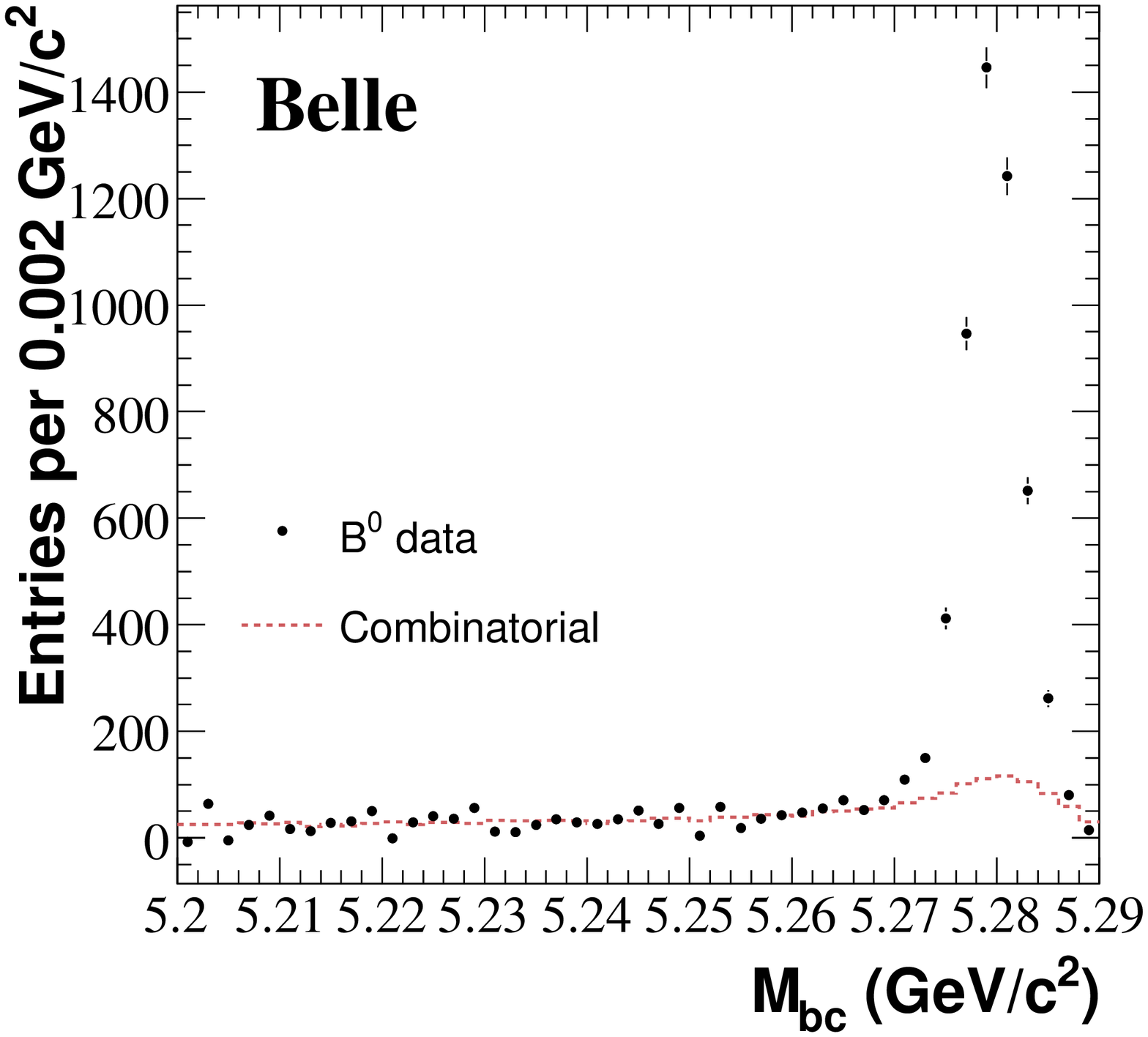}\\
\end{tabular}
\caption{The data points represent the beam-energy constrained mass, $M_{\rm{bc}}$, after electron selection cuts, $\Delta E$ cuts and continuum subtraction for the $B^+$ electron sample (left), and the $B^0$ electron sample (right).  The histogram represents the peaking combinatorial background determined from MC. }
\label{mbc}
\end{figure}

\subsubsection*{Combinatorial}
True $B\overline{B}$ events for which reconstruction or flavor assignment of the tagged $B$ meson is not correct are considered background events (which we refer to as combinatorial background). This background has a peaking structure in the signal region of $M_{\rm{bc}}$. 
We derive the shape of this background from generic $B \overline B$ MC events, where each particle used in the reconstruction of $B_{\rm tag}$ corresponds directly to what was generated in the simulation.
The yield of this  background is normalized to the on$-$resonance data $M_{\mathrm{bc}}$ sideband ($5.20~\mathrm{GeV/}c^2<M_{\mathrm{bc}}<5.25~\mathrm{GeV/}c^2$) after the subtraction of non-$B \overline{B}$ backgrounds.    Figure \ref{mbc} displays  the $M_{\rm bc}$ distribution after electron selection cuts, $\Delta E$ cuts and continuum subtraction, independently for  the $B^+$ and $B^0$ electron samples.  The contributions from the combinatorial background are overlaid.

\subsubsection*{Subtraction of $B \to X_u e \nu$}
The contribution of electrons from the inclusive $b \to u$ transition
are subtracted from the electron momentum spectrum.  This component of the background is normalized to the number of $B^+$ and $B^0$ tags, assuming the world average value for the inclusive charmless semileptonic branching fraction.

\subsubsection*{Fit to the inclusive spectra}
All remaining backgrounds arise when the fully reconstructed $B$ is correctly tagged, but the electron candidate either is from a secondary decay or is a misidentified hadron. These
background sources are irreducible. 

The background from $B \to D^{(*)} \to e$  decays is determined from MC simulation, adjusting the contribution of these events to the world average $B \to D$ {\it{anything}} and semileptonic $D$ branching fractions~\cite{PDG}.  Contributions from $J/ \psi$ and $\psi (2S)$ decays, photon conversions, and Dalitz decays, also determined by MC simulation, are small after our selection cuts.  Hadronic $B$ decays additionally contribute via hadron misidentification (i.e. $\pi$ fakes).

\begin{figure}[htb!]
  \begin{tabular}{cc}
    \includegraphics[width=0.48\textwidth]{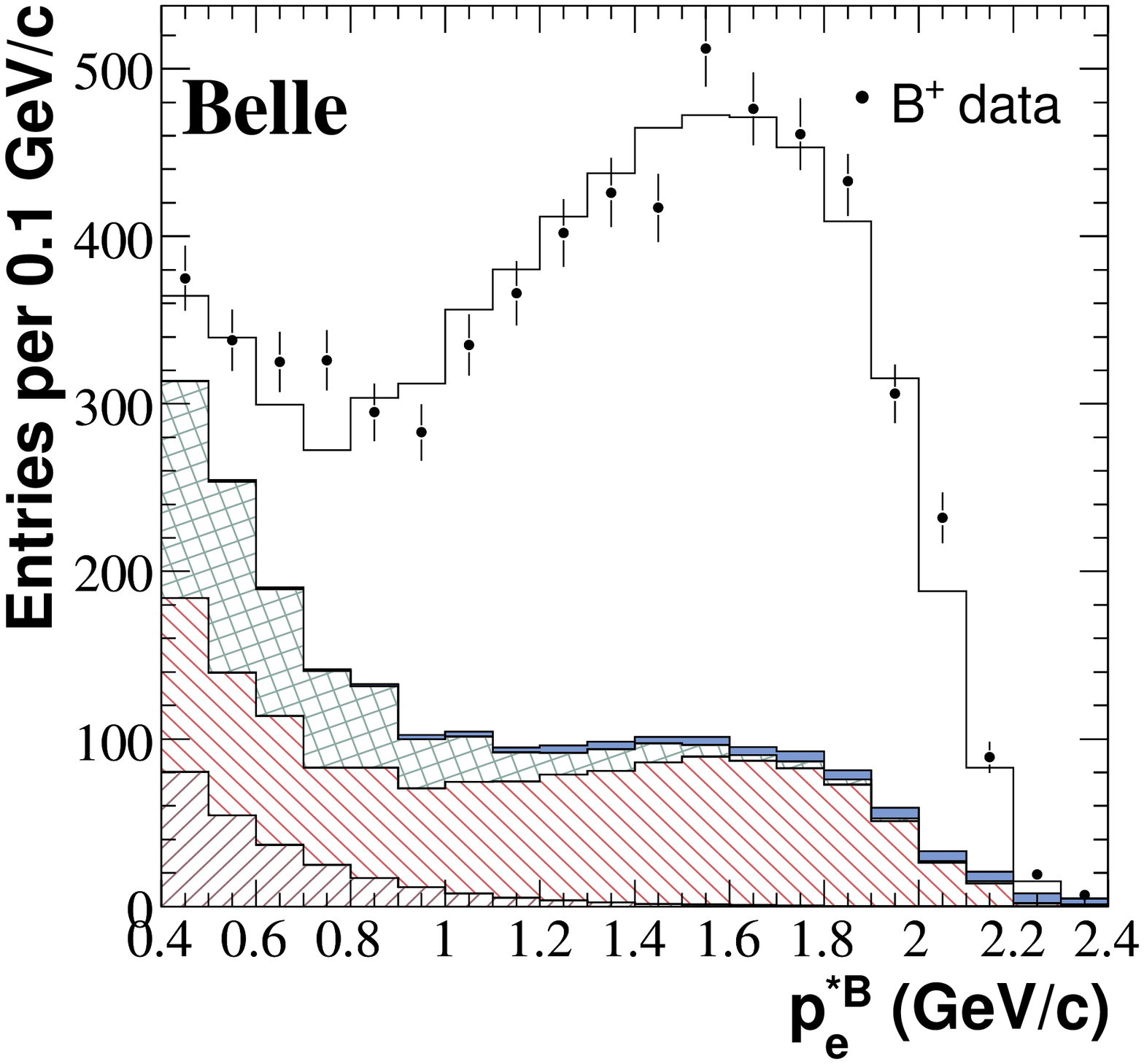}
    \includegraphics[width=0.48\textwidth]{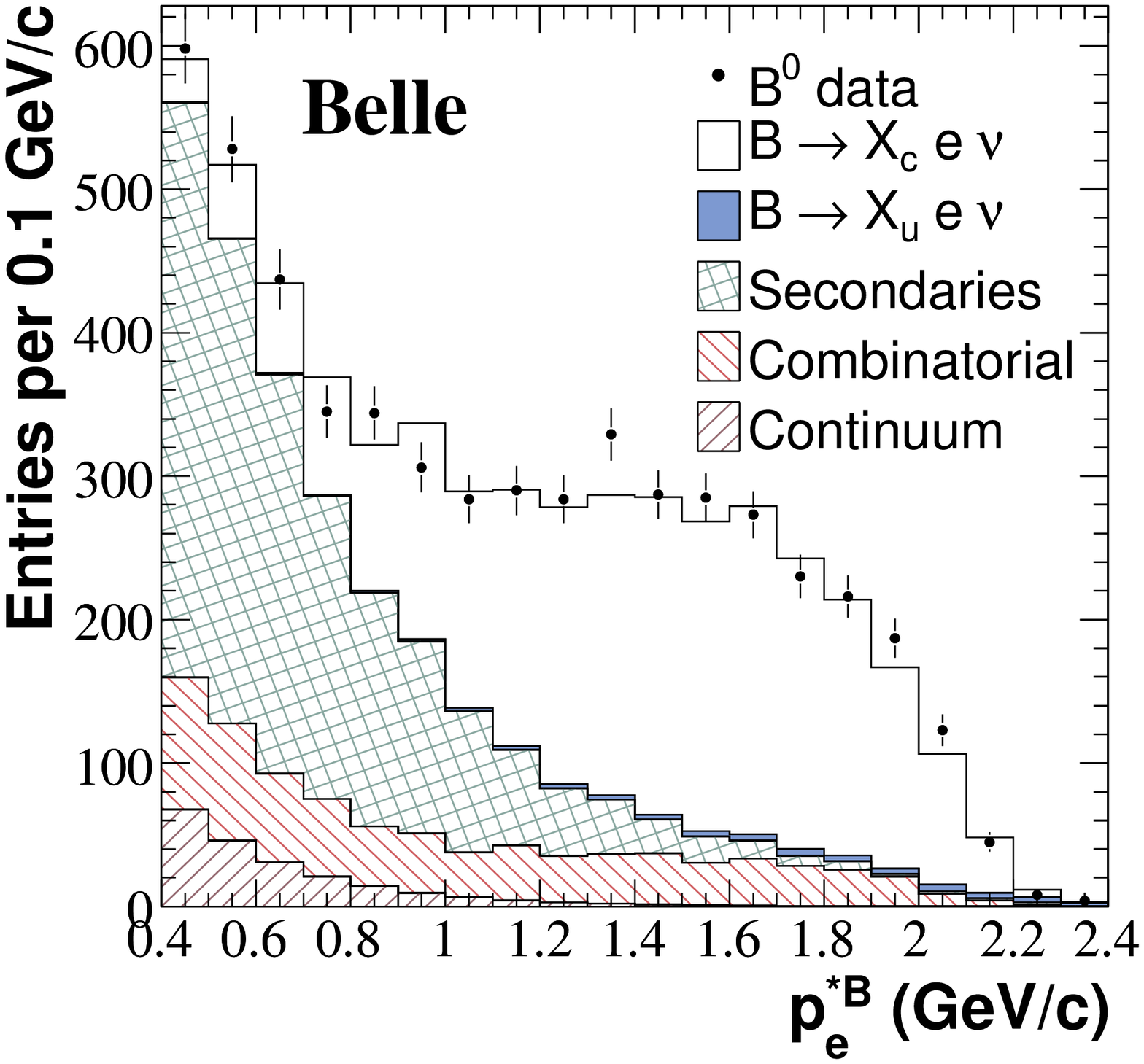} & \\
\end{tabular}
\caption{Measured electron momentum spectra from $B^+$ and $B^0$ decays before background subtraction, overlaid with the various backgrounds and the MC signal. Secondaries also includes hadron fakes. The errors shown are statistical only. }
\label{rawwithbg}
\end{figure}

We estimate the overall normalization of these remaining backgrounds by fitting the observed inclusive electron  momentum spectrum to the sum of the MC simulated signal and background contributions, after continuum, combinatorial and $B \to X_u e \nu$ background subtraction.  
The fit is performed in the range 0.4 GeV/$c$ $ < p_e^{*B} < $ 2.4 GeV/$c$, treating the relative normalization factors of the signal and background as the two free parameters in the fit.  The values of the $\chi^2$ per degree of freedom for the fits to $B^+$ and $B^0$ decay spectra are 1.3 and 1.1 respectively.  Figure \ref{rawwithbg} shows the electron momentum spectrum with
all background contributions overlaid, before corrections due to detector effects and selection efficiencies.    Confirmation of the agreement between the data and the signal and background MC can be seen in these plots, and has been furthermore checked in $M_{\rm bc}$ and $\Delta E$ sideband regions, where the signal contribution is less dominant.  The electron yields after particle selection cuts and subtraction of backgrounds are given in Table \ref{bzero}.

\begin{table}[htb]
\caption{Electron yields for $p^{*B}_{e} \geq 0.4$ GeV/$c$.  The errors are statistical only.}
\label{bzero}
\begin{center}
\begin{tabular}
{@{\hspace{0.5cm}}l|@{\hspace{0.5cm}}r@{{\hspace{0.15cm}}$\pm${\hspace{0.15cm}}}l@{\hspace{0.5cm}}@{\hspace{0.5cm}}r@{{\hspace{0.15cm}}$\pm${\hspace{0.15cm}}}l@{\hspace{0.5cm}}@{\hspace{0.5cm}}r@{{\hspace{0.15cm}}$\pm${\hspace{0.15cm}}}}

\hline \hline
$B$ candidate              & \multicolumn{2}{c}{$B^+$} \hspace{0.2cm} & \multicolumn{2}{c}{$B^0$} \\  
\hline\hline
   On Resonance Data          &  6423&    80&  5403&    74\\ 
\hline
   Scaled Off Resonance       &   249&    48&   209&    39\\ 
   Combinatorial Background&   1244&    20&   696&    13\\ 
\hline
   Secondary (Inc. Hadron Fakes)                           &   555&    11&  1843&    22\\ 
   $B \to X_u e \nu$               & 74     & 5     &57&6\\ 
\hline
   Background Subtracted    &  4300&    96     &  2597&    87\\ 
\hline \hline
\end{tabular}
\end{center}
\end{table}

\section{The Electron Energy Spectrum}
\subsection{Unfolding}
To measure the moments of the electron energy spectrum, we need to determine the true electron energy spectrum in the $B$ meson rest frame, $E_{e}^{*B}$.   In this analysis we assume the electron to be massless, imposing $E_{e}^{*B}=p_{e}^{*B}$.    The measured electron energy spectrum is distorted by various detector effects.
Hence, the true electron energy spectrum is extracted by performing an unfolding procedure based on 
the Singular Value Decomposition (SVD) algorithm~\cite{ref:13}.  The reliability of the unfolding procedure is dependent on the agreement between data and MC simulation, both for the physics models and the detector response.  Studies of MC show that there are no biases due to the SVD unfolding algorithm.

\begin{figure}[htb]
    \includegraphics[width=0.48\textwidth]{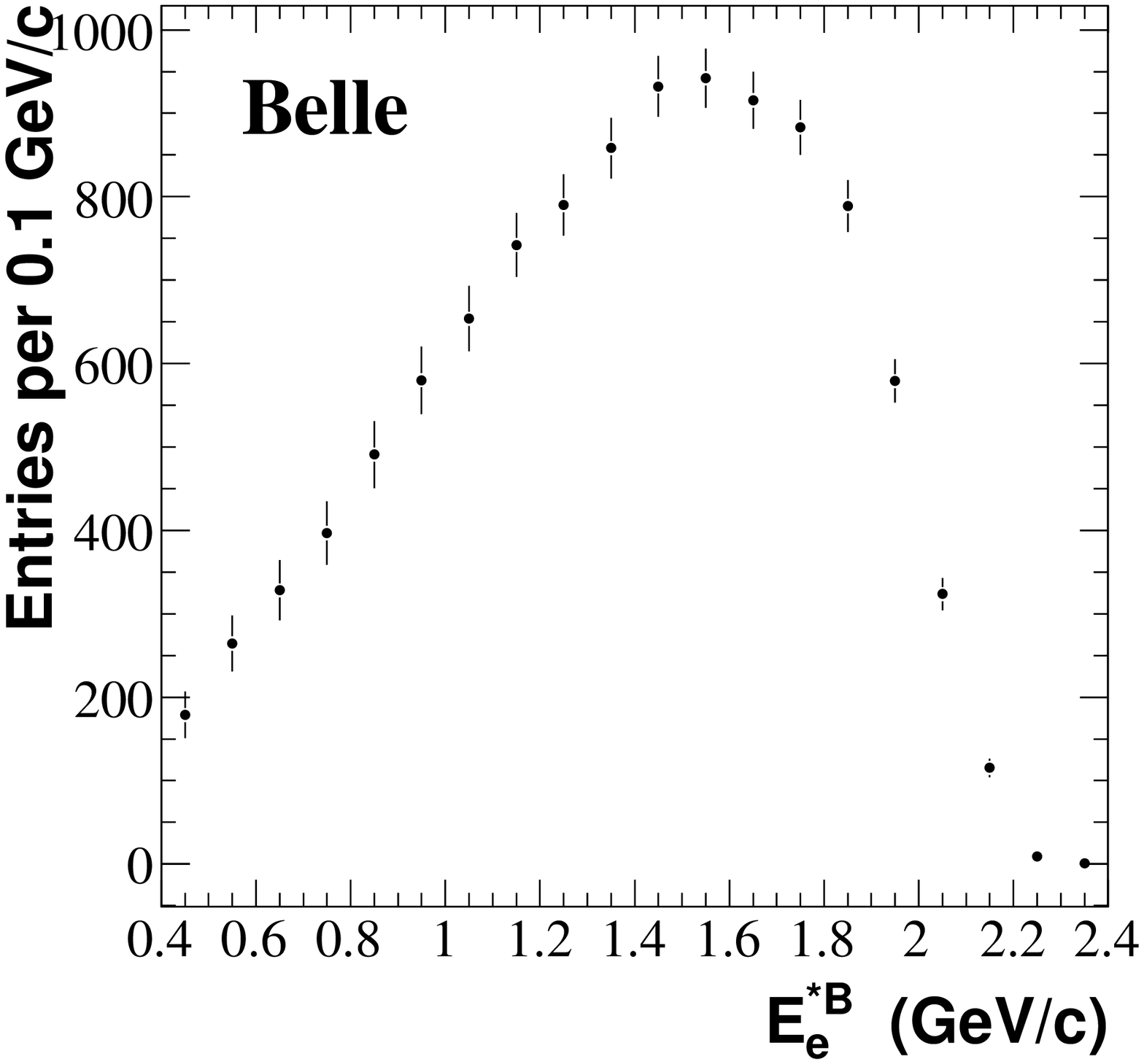}
    \includegraphics[width=0.48\textwidth]{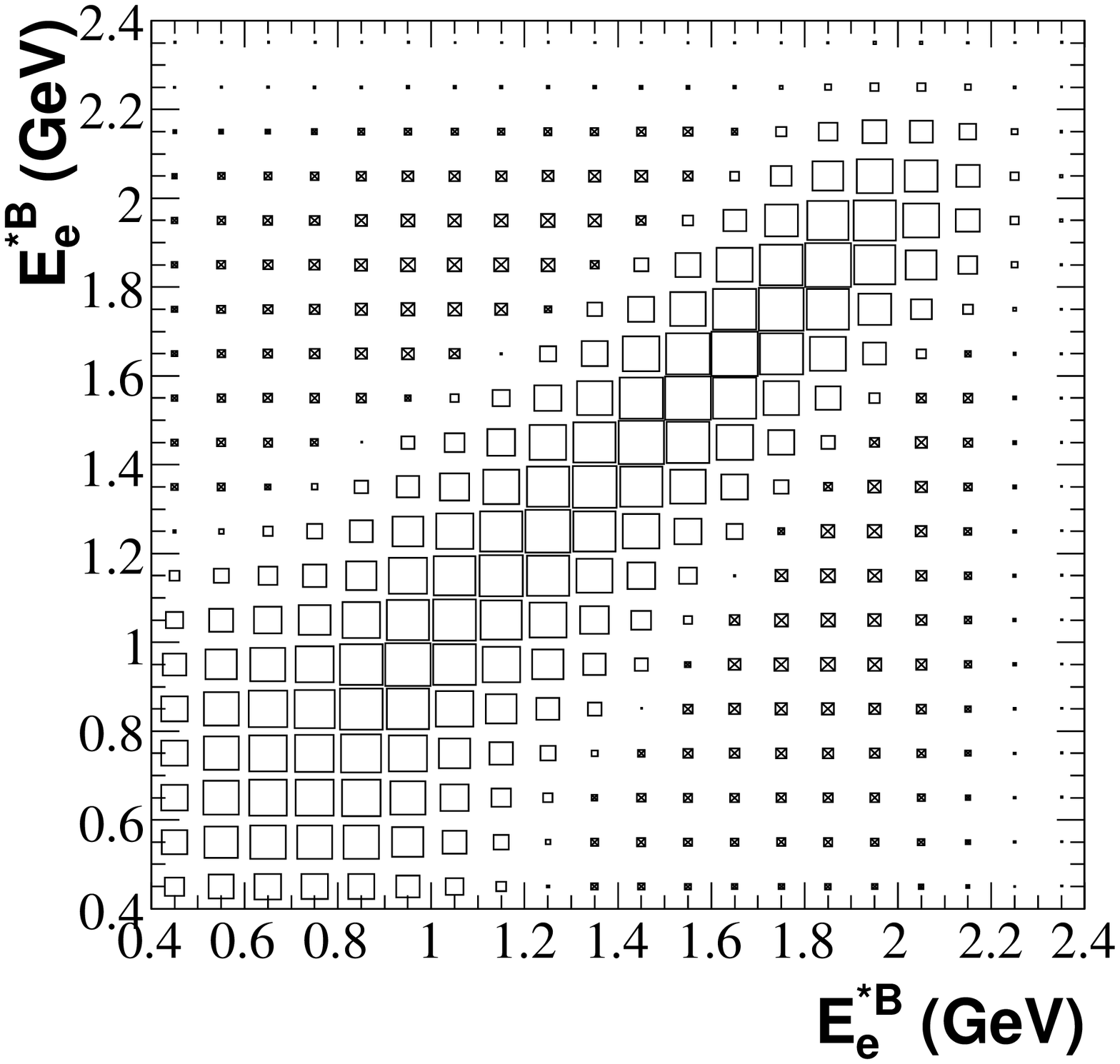}
  \caption{Unfolded electron energy distribution in the $B$ meson rest frame(left), combining contributions from $B^0$ and $B^+$ decays, and corrected for QED radiative effects, detector and selection efficiencies. The errors shown are statistical. On the right is the corresponding unfolded electron energy distribution covariance matrix, where the filled boxes represent negative elements. (These plots are shown for illustrative purposes only.)}
  \label{covariance}
\end{figure}

 The unfolded spectrum is corrected for QED radiative effects using the PHOTOS algorithm~\cite{PHOTOS}, as the OPE does not have $\mathcal{O}(\alpha)$ QED corrections.  The unfolded electron energy spectrum and the bin-to-bin statistical covariance matrix calculated with the unfolding algorithm are  shown in Fig. \ref{covariance} (for illustrative purposes only,  as the full error analysis is performed on a moment measurement basis).
 
\subsection{Moments and Partial Branching Fractions}

We measure the first four central moments of the electron energy spectrum with nine electron energy threshold values
($E_{\mathrm{cut}}=$ 0.4, 0.6, 0.8, 1.0, 1.2, 1.4, 1.6, 1.8, 2.0 GeV in the $B$ rest frame) combining the spectra from $B^+$ and $B^0$ semileptonic decays. The first  moment is defined to be
$M_1^I = \langle E^{*B}_e \rangle _{E^{*B}_e>E^I_{\rm cut}}$ and subsequent central moments are calculated about the
first moment,
$M_n^I= \langle (E^{*B}_e-  M_1^I )^n \rangle _{E^{*B}_e>E^I_{\rm cut}}$,  where $I$ is the index for the electron energy threshold and $n=2,3,4$.  The statistical uncertainty of each moment is calculated as:
\begin{equation}
\sigma_{\rm{stat}}^2(M^I_n)=\frac{\Sigma_{ij}(E^{*B}_{e,i}- M^I_1  )^n X_{ij}(E^{*B}_{e,j}- M^I_1  )^n}{(\Sigma_i x'_i)^2} _{E^{*B}_{e,i(j)}>E_{\rm cut}^I},
\end{equation}
where $x'$ is the unfolded spectrum in the $B$ rest frame, corrected for the bin-to-bin detection efficiencies, $X$ is the covariance matrix, and $E^{*B}_{e,i(j)}$ is the central value of bin $i (j)$ in the $B$ rest frame.  

In addition, we measure the partial branching fractions, combining the spectra from $B^+$ and $B^0$ semileptonic decays, and independently measuring the  $B^+$ and $B^0$  partial branching fractions at the previous lower electron energy threshold of 0.6 GeV \cite{okabe} and a new lower electron energy threshold of 0.4 GeV.  The expression for the partial branching fraction and its statistical error is calculated as:
\begin{equation}
\Delta\mathcal{B}^I= \frac{\Sigma_i (x''_i)}{N_{\rm{tag}}} _{E^{*B}_{e,i(j)}>E_{\rm cut}^I},~\sigma^2_{\rm{stat},\Delta\mathcal{B}^I}= \left[ (\frac{\sqrt{\Sigma_{ij} X''_{ij}}}{{N_{\mathrm{tag}}}})^2 + (\frac{\Sigma_i (x''_i) \Delta N_{\rm{tag}} }{N_{\rm{tag}}^2})^2 \right] _{E^{*B}_{e,i(j)}>E_{\rm cut}^I},
\end{equation}
where $N_{\rm{tag}}$ is the number of tagged $B$ events, and $x''$ and $X''$ denote the full efficiency corrected unfolded spectrum and covariance matrix respectively.

\section{Systematic Errors}
The contributions to the systematic error  for each moment and electron energy threshold are summarized in Tables \ref{sys1}, \ref{sys2}, \ref{sys3}, \ref{sys4}, \ref{sys0} for all moments and partial branching fraction measurements.   The total systematic error is obtained by adding each contribution in quadrature.  The principal systematic errors originate from the event selection, electron identification, background estimation and signal model dependence.

\subsection{Detector related uncertainties}
The selection efficiency for $B \to X_c e \nu$ decays is determined by MC simulation.  There are three major factors that determine the detection efficiency: the track reconstruction of the electron, the electron identification, and event selection.

The uncertainty in the tracking efficiency has been studied in detail in Ref.~\cite{okabe}, which is estimated to be a 1\% effect on the overall efficiency.  The electron identification efficiency is determined with a radiative Bhabha sample with dependence on the electron energy in the laboratory frame, and the angle subtended by the electron in the detector.  The effect of the difference between the $B \overline B$ event environment and the simpler radiative Bhabha environment (two charged tracks and one shower) is studied with embedded samples and a correction for this bias is performed on the measured spectrum.  This bias decreases at higher electron momenta.  The systematic error associated to the difference between MC and data tracking resolution is negligible.  We assess the impact of these uncertainties on the observed spectrum for both the signal and the background.  Improvements have been made in the understanding of this systematic error with respect to similar previous measurements~\cite{okabe}.
 
\subsection{Uncertainties in the signal spectrum}
The branching fractions for exclusive semileptonic $B \to X_c e \nu$ are not precisely known, particularly $D^{**}$ contributions. For this reason, we introduce a scale factor in the background fits to adjust the overall normalization of the prompt contribution.  In addition, we adjust the individual branching fractions of each exclusive $B \to X_c e \nu$ decay mode.  To test the sensitivity to the shape of the signal contributions, we have varied the form factors for the prompt decay types $D^* e \nu$, and $D e \nu$, and changed the model input parameters which describe the differential decay rates of the resonant $D^{**} e \nu$ decays.

For $B \to D^{(*)} e \nu$ decays we use HQET parameterizations of the form factors.  To study the impact of the uncertainties in the measured form factors, we reweight the MC-simulated spectrum for a given decay mode to reproduce the change in the spectrum due to the variations of the form-factor parameters.  From the observed changes in the  signal yield and shape, as a function of the choice of the form factor parameters for $D^* e \nu$ decays, we assess the systematic error on the moments by varying the form-factor parameters, $\rho^2$ \cite{HFAG}, $R_1$ and $R_2$ by one standard deviation \cite{dstarbabar}.  For $D e \nu$ decays, we rely on measurements of $\rho^2_D$.  Similarly, we estimate the impact of the uncertainty in $\rho^2_D$  \cite{HFAG} by comparing the change on the moments, corresponding to variations of $\rho^2_D$ by one standard deviation.  

To assess the impact of the poorly known branching fractions and differential decay rates for various resonant $D^{**} e \nu$ decay we take into account limits from measurements to resonant and non-resonant $D^{(*)} \pi e \nu$ states, and full inclusive rates~\cite{matsumoto,kuzmin,LEP,PDG}.  We determine the systematic variation on the moments by varying the LLSW~\cite{LLSW} model parameters for the differential decay rates, within their allowed ranges imposed by measurement and theory arguments.  Predictions for $D^{**} e \nu$ shapes and branching fractions are assumed to be fully correlated as they rely on the same set of  parameters. We use half of  the shift between the LLSW model parameter bounds, as an estimate of the systematic error due to the uncertainty in $B$ to $D^{**}$ decays.
The branching fraction for non-resonant $D^{(*)} \pi e \nu$ decay modes are assumed to be uncorrelated with the $D^
{**}$ decays, and the systematic variation on the moments is estimated as half of the shift between the bounds on the branching fraction.

\subsection{Background subtraction}
Systematic errors in the subtraction of the non-$B \overline{B}$ background are dominated by the uncertainty in the relative normalization of the on- and off-resonance data.
The error arises from the uncertainty of the measured luminosities, which is estimated to be a 1\% error on the continuum electron yield.

The shapes of the $B \overline{B}$ backgrounds are derived from MC simulations.  The uncertainty due to mis-tagging in the $B^0$ and $B^+$ samples is estimated by varying the lower bound on the $M_{\rm{bc}}$ signal region, corresponding to a 10\% variation in the ratio of good tags to incorrect tags in the signal region.

The uncertainty due to the $b \to u$ subtraction, which occurs before  unfolding, is evaluated by varying the total inclusive charmless branching fraction by one standard deviation~\cite{HFAG}. 

The uncertainty due to secondary, cascade $B \to D \to e$ decays is assessed by varying the branching fractions of semileptonic $D$ decays, and $B \to D$ {\it anything} by one standard deviation~\cite{PDG}.  This contribution is significant in the neutral $B$ sample as there is no cut on the correlation between the flavor of the tagging $B$ and the electron charge.
For background from hadronic $B$ decays, the uncertainty is primarily due to the uncertainty in the hadron misidentification.  The uncertainty associated to the magnitude of the hadron fake contribution is determined from a comparison of the fake rates measured with $K^0_S \to \pi^+ \pi^-$ decays in real data and in the MC simulation.

The systematic uncertainty due to the overall fit for the secondaries to the data is estimated by varying the lower $p_e^{*B}$ bound of the fit region, and the number of bins used in the fit.

\subsection{Unfolding and Radiative Corrections}
For the uncertainties related to the unfolding procedure, we vary the effective rank parameter by one in the SVD algorithm.  Corrections for QED radiation in the decay process are simulated using PHOTOS.  The simulation includes multiple-photon emission from the electron, and interference effects.  The accuracy of this simulation has been compared to analytical calculations performed at ${\it{O}}(\alpha)$~\cite{elzbieta}.  Based on this comparison, the uncertainty of the PHOTOS correction leads to a negligible contribution to the overall systematic error.

\section{Correlations}
All measurements performed on the $B \to X_c e \nu$ electron energy spectrum are correlated, with both overlapping data samples, and common systematics.  The following describes the procedure for calculating the covariance and correlations between measurements of the partial inclusive branching fractions and the moments at varying threshold energies.

The statistical covariance matrix of two correlated moment measurements, $\mathrm{cov_{stat}}[M_k^I,M_l^J]$, is simply a general case of the error calculation:
\begin{equation}
\mathrm{cov_{stat}}[M^{I}_{k},M^{J}_{l}]=\frac{\Sigma_{ij}(E^{*B}_{e,i}- M^I_1 )^k X_{ij}(E^{*B}_{e,j}-M^J_1  )^l}{\Sigma_i (x'_i)\Sigma_j (x'_j)} _{E^{*B}_{e,i}>E_{\rm cut}^I,~E^{*B}_{e,j}>E_{\rm cut}^J},
\label{covstat}
\end{equation}
where $k$ and $l$ are the order of the moments and $I$ and $J$ are indices for the threshold energies.
For covariance matrices including the partial branching fractions (zeroth moments), $\Delta\mathcal{B}_{I(J)}$, the factors  $1/\Sigma_{i(j)} (x^{'}_{i(j)})$ in equation \ref{covstat} are not present.

The systematic covariance matrix is calculated assuming correlations between individual systematic variations to be positive ($\rho=1$), negative ($\rho=-1$) or zero ($\rho=0$), thus;

\begin{equation}
\label{cor1}
\mathrm{cov_{sys}}[M^{I}_{k},M^{J}_{l}]=\mathrm{\rho}_{{\rm{sys}} ( M_k ^I ,  M_l ^J)}\sigma_{{\rm{sys}},  M_k ^I}\sigma_{{\rm{sys}},  M_l ^J}.
\label{coreq}
\end{equation}
 summing over all systematic variations.

To obtain the overall covariance matrix, we add the statistical and systematic covariance matrices together.  The total correlation between measurements, $\rho_{M_k^I,M_l^J}$, is then derived from the overall covariance matrix and the total errors for each measurement, using a similar expression to equation \ref{coreq}.  

\section{Results}
\begin{figure}[htb]
  \begin{tabular}{cc}
  \includegraphics[width=0.43\textwidth]{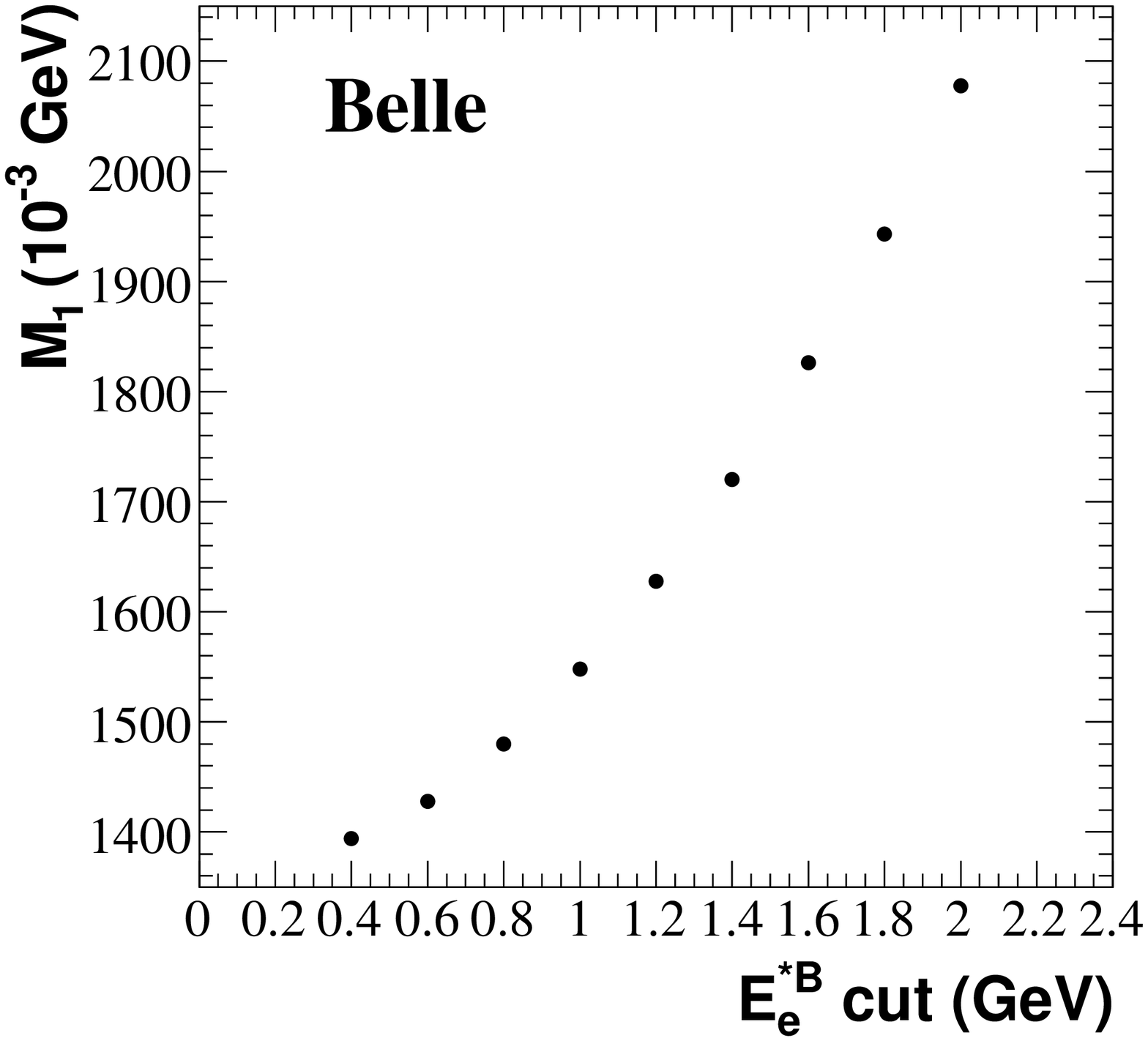}&
  \includegraphics[width=0.43\textwidth]{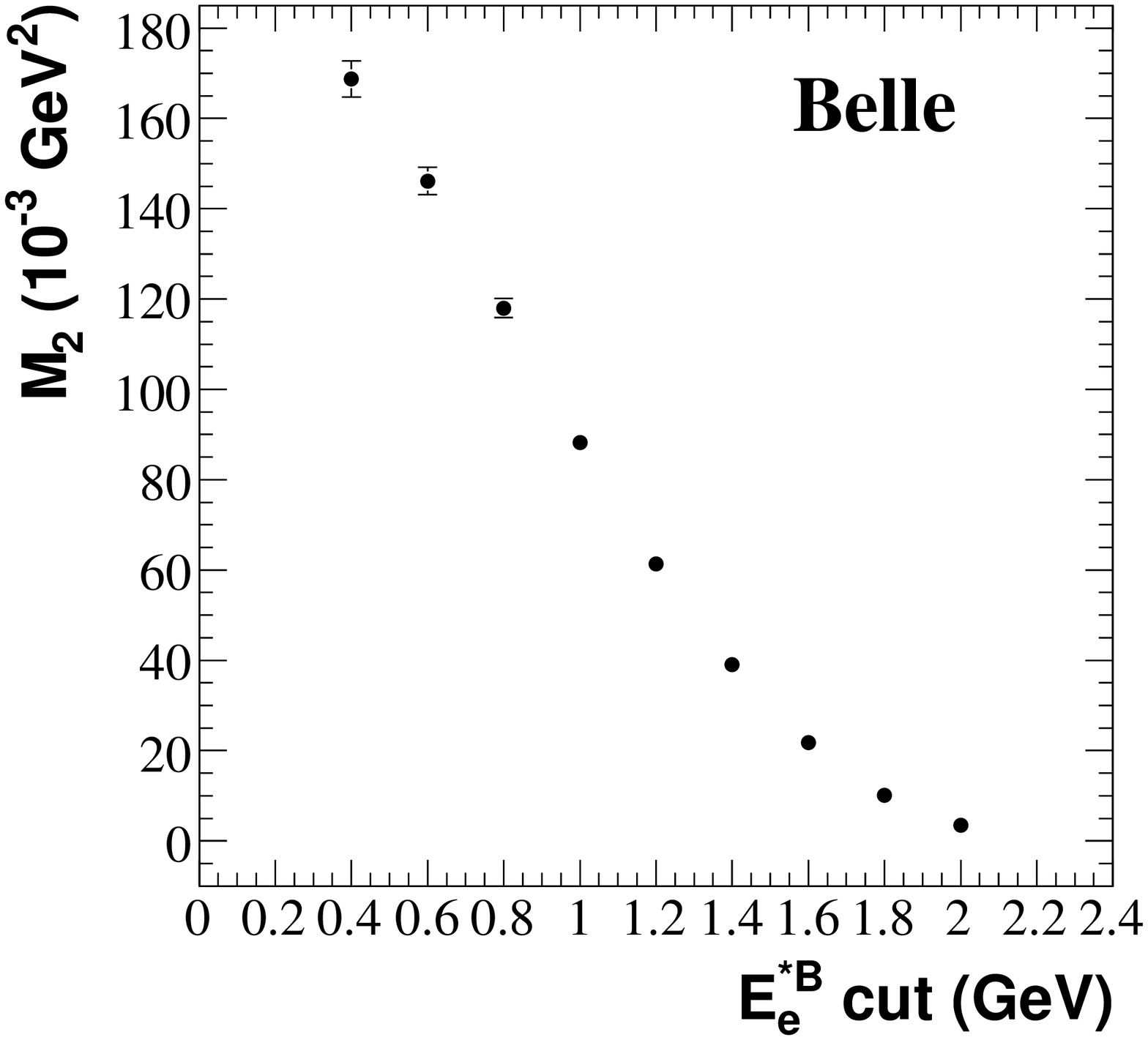}\\
  \includegraphics[width=0.43\textwidth]{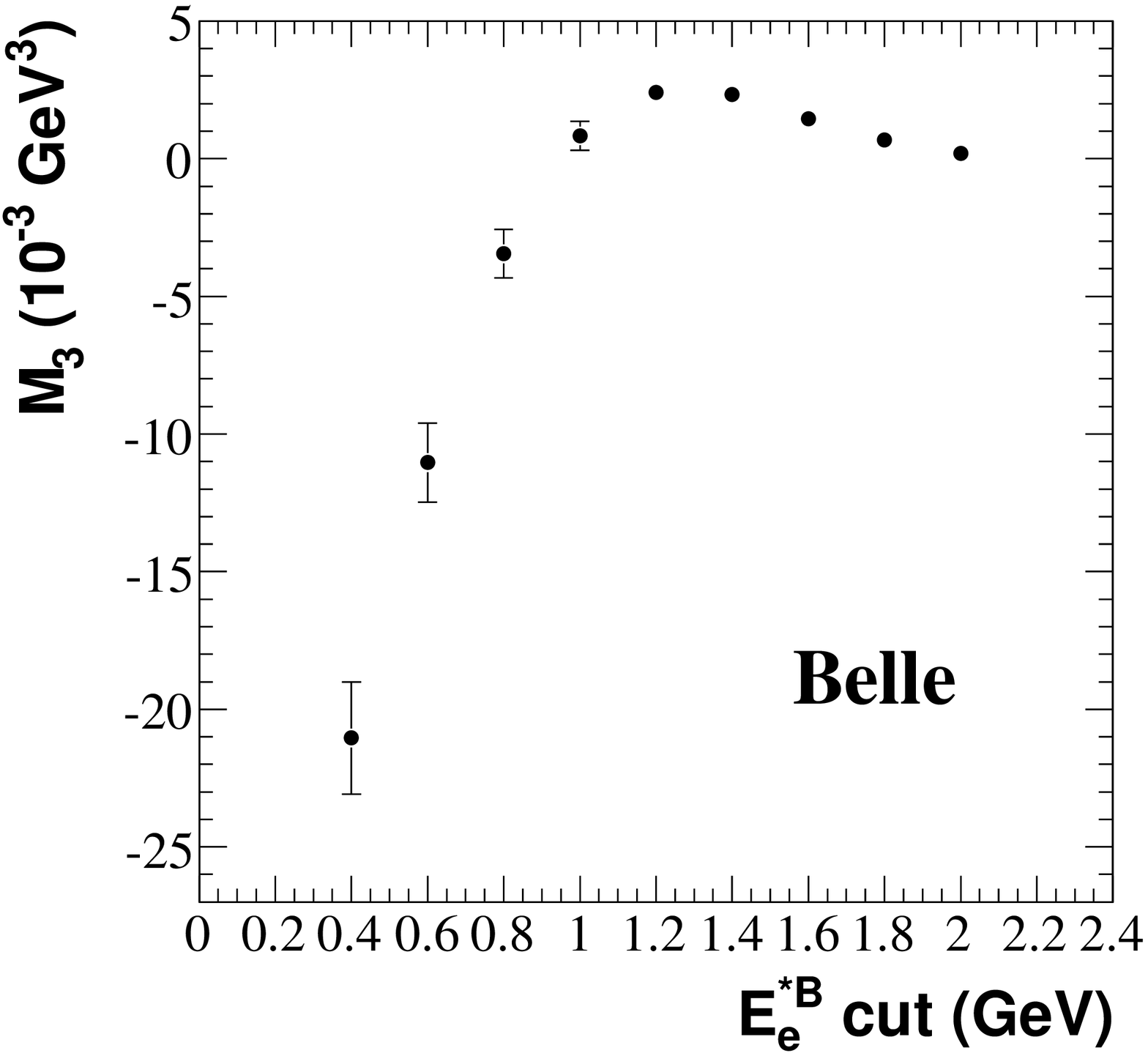}&
  \includegraphics[width=0.43\textwidth]{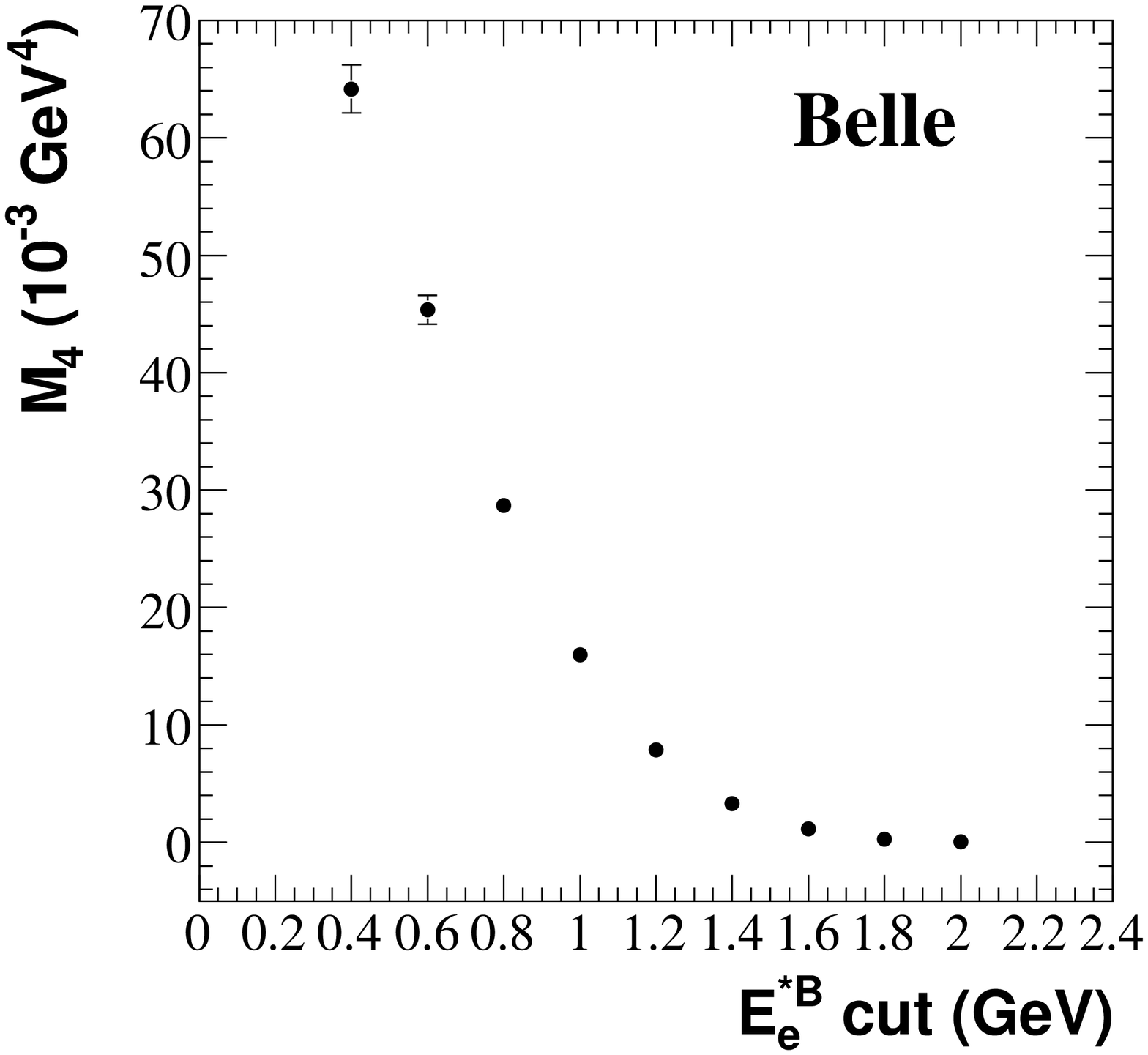}\\
  \includegraphics[width=0.43\textwidth]{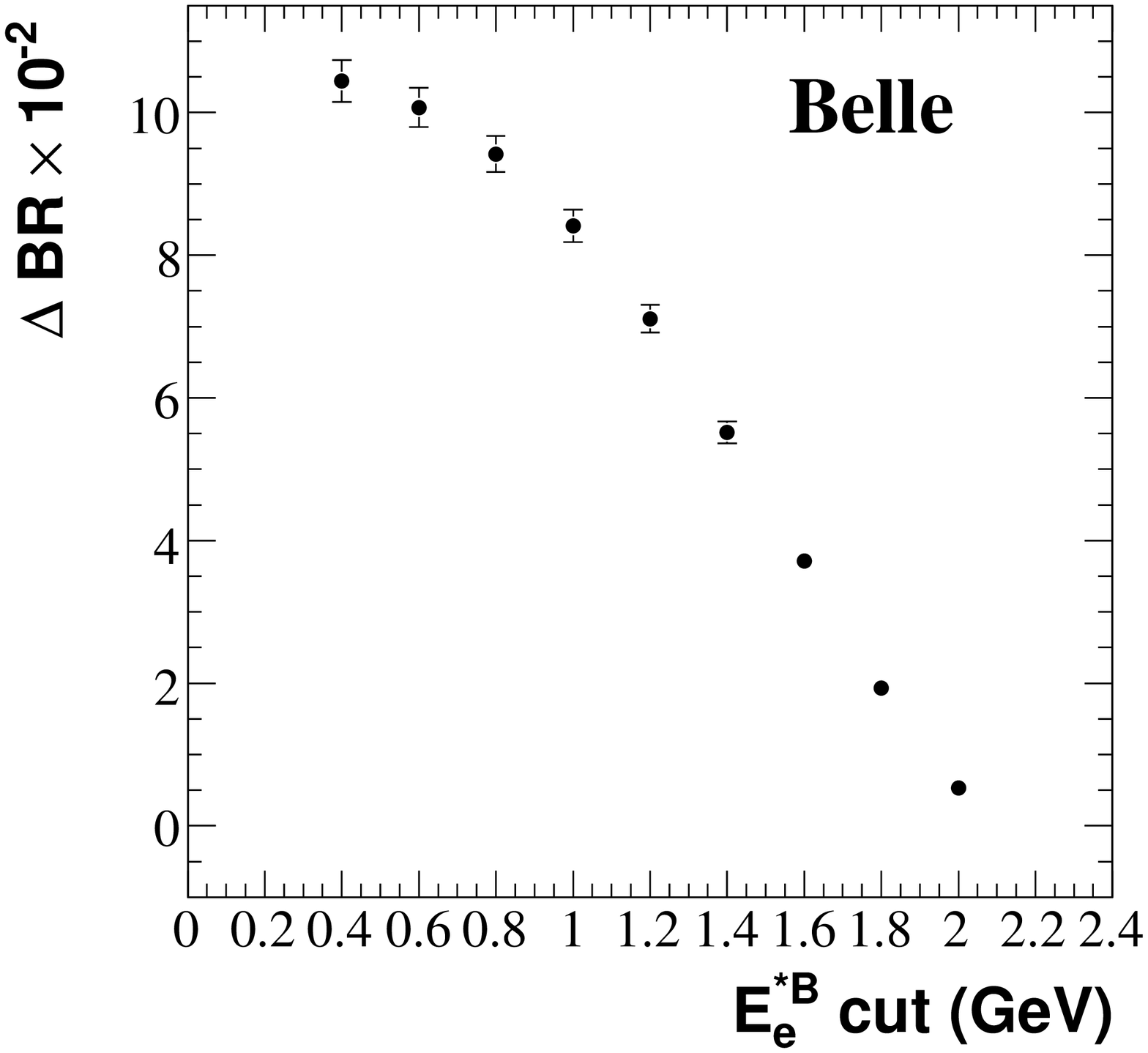}\\
  \end{tabular}
\caption{First, second, third and fourth electron energy moments and partial branching fractions ($M_1$, $M_2$, $M_3$, $M_4$, $\Delta\mathcal{B}$), 
as a function of the electron threshold energy $E_{\mathrm{cut}}$. The errors shown are the statistical and
  systematic errors added in quadrature.}
\label{momentplots}
\end{figure}

Table \ref{moments2} provides the  $B^0$ and $B^+$ weighted average moments as a function of $E_{\mathrm{cut}}$.  Figure \ref{momentplots} illustrates these results.  The measurement of the first electron energy moment, $M_1$, at $E_{\rm cut}=0.6$ GeV, is $(1427.82 \pm  5.82({\rm stat.}) \pm  2.55({\rm sys.}))$ MeV, which is consistent with, and improves upon measurements by BABAR~\cite{babarel} and CLEO~\cite{cleoel}.  
The independent partial branching fraction measurements of $B^+$ and $B^0$ at 0.4 GeV and 0.6 GeV electron energy thresholds are provided with a breakdown of their systematic uncertainties in Table \ref{bplusbzerosys}.  The results, $\Delta \mathcal{B} (B^+ \to X_c e \nu$, $E_{\rm{cut}} = 0.6 ~{\rm GeV})=(10.34 \pm 0.23({\rm stat.}) \pm 0.25({\rm sys.}))$\% and $\Delta \mathcal{B} (B^0 \to X_c e \nu$, $E_{\rm{cut}} = 0.6 ~{\rm GeV})=(9.80 \pm 0.29({\rm stat.}) \pm 0.21(\rm{ sys.}))$\%, are consistent with our previous measurements~\cite{okabe}, with the overall errors improved by approximately 30\%.  The observed $\Delta \mathcal{B} (B^+ \to X_c e \nu)$/$\Delta \mathcal{B} (B^0 \to X_c e \nu)$ ratio, at $E_{\rm cut}=$0.4 GeV, is $1.07 \pm 0.04({\rm stat.}) \pm 0.03({\rm sys.})$, is consistent with the $B^+/B^0$ lifetime ratio at $\tau_+/\tau_0=1.076 \pm 0.008$~\cite{PDG}. 
The correlation coefficients for each moment (including the averaged partial branching fractions) and cut combination, are presented in Tables \ref{m1m1} through to \ref{m0m0}.

\newcolumntype{Z}{>{\raggedleft\arraybackslash}X}

\section{Summary}
We report a measurement of the electron energy spectrum of the
inclusive decay $B \rightarrow X_c e \nu$ and its first four moments for threshold energies from 0.4 GeV to 2.0 GeV.  In addition we provide the partial branching fraction measurements for the same set of threshold energies, including independent measurements of $B^+$ and $B^0$ at threshold energies of 0.4 GeV and 0.6 GeV. The full correlation matrix for this set of measurements has also been evaluated.  This set of moments, combined with the  moments of the hadronic  mass distribution, can serve as input for the determination of HQE parameters and of $|V_{cb}|$.

\begin{table}[htb]
\caption{Breakdown of the systematic errors for the first moment, $M_1$, for $B \rightarrow X_c e \nu$ in the $B$ meson rest frame for nine values of the electron energy threshold $E_{\mathrm{cut}}$.}
\label{sys1}
\begin{center}
\begin{tabularx}{0.9\linewidth}{l|ZZZZZZZZZ}
\hline \hline
                            & \multicolumn{9}{c}{$M_1$  [$\rm MeV$]}\\
$E_{\mathrm{cut}}$[GeV]          &0.4&0.6&0.8&1.0&1.2&1.4&1.6&1.8&2.0\\
\hline\hline
                                                                   Electron Detection&     0.81&     0.77&     0.42&     0.16&     0.05&     0.09&     0.02&     0.02&     0.01\\
 \hline
                                                                $(D^{(*)} e \nu)$ form factors&     0.59&     0.62&     0.61&     0.47&     0.33&     0.75&     0.80&     0.99&     1.20\\
                                                         $\mathcal{B}(D^{(*)} e \nu)$&     0.22&     0.17&     0.11&     0.11&     0.09&     0.09&     0.05&     0.10&     0.07\\
                                                                  $(D^{**}e\nu)$ form factors&     1.71&     1.03&     0.47&     0.10&     0.09&     0.08&     0.10&     0.12&     0.02\\
   $\mathcal{B} (D^{(*)}_{\rm{non-res}}                   \pi e \nu / D^{**}  e \nu)$&     1.15&     1.37&     1.50&     0.96&     0.66&     0.38&     0.28&     0.30&     0.16\\
 \hline
                                                                            Continuum&     0.02&     0.00&     0.02&     0.02&     0.01&     0.01&     0.01&     0.00&     0.00\\
                                                                        $M_{\rm{bc}}$&     1.14&     0.72&     0.24&     0.02&     0.03&     0.05&     0.08&     0.05&     0.04\\
                                                                          $X_u e \nu$&     0.79&     0.78&     0.77&     0.75&     0.72&     0.67&     0.56&     0.36&     0.14\\
                                                                         Hadron Fakes&     0.65&     0.56&     0.42&     0.24&     0.11&     0.04&     0.01&     0.00&     0.00\\
                                                                $B \to D^{(*)} \to e$&     0.91&     0.79&     0.60&     0.39&     0.21&     0.10&     0.04&     0.01&     0.00\\
                                                                          Secondaries&     0.82&     0.68&     0.49&     0.27&     0.12&     0.04&     0.01&     0.00&     0.00\\
 \hline
                                                                            Unfolding&     0.02&     0.23&     0.28&     0.24&     0.04&     0.06&     0.10&     0.33&     0.04\\
 \hline
                                                                    Total Systematics&     3.02&     2.55&     2.13&     1.45&     1.08&     1.10&     1.03&     1.16&     1.23\\
\hline\hline
\end{tabularx}
\end{center}
\end{table}

\begin{table}[htb]
\caption{Breakdown of the systematic errors for the second moment, $M_2$, for $B \rightarrow X_c e \nu$ in the $B$ meson rest frame for nine values of the electron energy threshold  $E_{\mathrm{cut}}$.}
\label{sys2}
\begin{center}
\begin{tabularx}{0.9\linewidth}{l|ZZZZZZZZZ}

\hline \hline
                            & \multicolumn{9}{c}{$M_2$  [$10^{-3}\rm GeV^2$]}\\
$E_{\mathrm{cut}}$[GeV]          &0.4&0.6&0.8&1.0&1.2&1.4&1.6&1.8&2.0\\
\hline\hline
                                                                   Electron Detection&     0.27&     0.31&     0.16&     0.08&     0.03&     0.01&     0.00&     0.00&     0.00\\
 \hline
                                                                $(D^{(*)} e \nu)$ form factors&     0.55&     0.38&     0.34&     0.27&     0.31&     0.21&     0.19&     0.18&     0.19\\
                                                         $\mathcal{B}(D^{(*)} e \nu)$&     0.12&     0.11&     0.02&     0.04&     0.02&     0.02&     0.02&     0.01&     0.02\\
                                                                  $(D^{**}e\nu)$ form factors&     0.87&     0.44&     0.18&     0.06&     0.04&     0.02&     0.02&     0.00&     0.00\\
   $\mathcal{B} (D^{(*)}_{\rm{non-res}}                   \pi e \nu / D^{**}  e \nu)$&     0.75&     0.66&     0.23&     0.08&     0.09&     0.04&     0.04&     0.01&     0.01\\
 \hline
                                                                            Continuum&     0.00&     0.00&     0.00&     0.00&     0.00&     0.00&     0.00&     0.00&     0.00\\
                                                                        $M_{\rm{bc}}$&     0.38&     0.24&     0.07&     0.03&     0.01&     0.01&     0.01&     0.00&     0.00\\
                                                                          $X_u e \nu$&     0.27&     0.25&     0.23&     0.19&     0.15&     0.11&     0.06&     0.03&     0.01\\
                                                                         Hadron Fakes&     0.15&     0.11&     0.06&     0.03&     0.01&     0.00&     0.00&     0.00&     0.00\\
                                                                $B \to D^{(*)} \to e$&     0.19&     0.13&     0.07&     0.03&     0.01&     0.00&     0.00&     0.00&     0.00\\
                                                                          Secondaries&     0.20&     0.14&     0.07&     0.03&     0.01&     0.00&     0.00&     0.00&     0.00\\
 \hline
                                                                            Unfolding&     0.56&     0.34&     0.04&     0.02&     0.00&     0.02&     0.07&     0.01&     0.00\\
 \hline
                                                                    Total Systematics&     1.53&     1.08&     0.55&     0.36&     0.36&     0.34&     0.22&     0.18&     0.19\\
                                                                    \hline\hline
\end{tabularx}
\end{center}
\end{table}

\begin{table}[htb]
\caption{Breakdown of the systematic errors for the third moment, $M_3$, for $B \rightarrow X_c e \nu$ in the $B$ meson rest frame for nine values of the electron energy threshold $E_{\mathrm{cut}}$.}
\label{sys3}
\begin{center}
\begin{tabularx}{0.9\linewidth}{l|ZZZZZZZZZ}

\hline \hline
                            & \multicolumn{9}{c}{$M_3$  [$10^{-3}\rm GeV^3$]}\\

$E_{\mathrm{cut}}$[GeV]          &0.4&0.6&0.8&1.0&1.2&1.4&1.6&1.8&2.0\\
\hline\hline
                                                                   Electron Detection&     0.15&     0.06&     0.03&     0.01&     0.01&     0.00&     0.00&     0.00&     0.00\\
 \hline
                                                                $(D^{(*)} e \nu)$ form factors&     0.26&     0.17&     0.15&     0.13&     0.09&     0.06&     0.05&     0.04&     0.03\\
                                                         $\mathcal{B}(D^{(*)} e \nu)$&     0.01&     0.03&     0.02&     0.01&     0.00&     0.01&     0.00&     0.00&     0.00\\
                                                                  $(D^{**}e\nu)$ form factors&     0.07&     0.03&     0.03&     0.01&     0.00&     0.00&     0.00&     0.00&     0.00\\
   $\mathcal{B} (D^{(*)}_{\rm{non-res}}                   \pi e \nu / D^{**}  e \nu)$&     0.46&     0.35&     0.18&     0.10&     0.04&     0.02&     0.01&     0.00&     0.00\\
 \hline
                                                                            Continuum&     0.00&     0.00&     0.00&     0.00&     0.00&     0.00&     0.00&     0.00&     0.00\\
                                                                        $M_{\rm{bc}}$&     0.16&     0.12&     0.08&     0.02&     0.02&     0.00&     0.00&     0.00&     0.00\\
                                                                          $X_u e \nu$&     0.07&     0.06&     0.04&     0.03&     0.02&     0.01&     0.00&     0.00&     0.00\\
                                                                         Hadron Fakes&     0.10&     0.08&     0.06&     0.03&     0.01&     0.00&     0.00&     0.00&     0.00\\
                                                                $B \to D^{(*)} \to e$&     0.10&     0.09&     0.07&     0.03&     0.01&     0.00&     0.00&     0.00&     0.00\\
                                                                          Secondaries&     0.08&     0.07&     0.05&     0.02&     0.01&     0.00&     0.00&     0.00&     0.00\\
 \hline
                                                                            Unfolding&     0.28&     0.20&     0.13&     0.10&     0.03&     0.03&     0.01&     0.00&     0.00\\
 \hline
                                                                    Total Systematics&     0.66&     0.49&     0.30&     0.20&     0.11&     0.07&     0.05&     0.04&     0.03\\
\hline\hline
\end{tabularx}
\end{center}
\end{table}

\begin{table}[htb]
\caption{Breakdown of the systematic errors for the fourth moment, $M_4$, for $B \rightarrow X_c e \nu$ in the $B$ meson rest frame for nine values of the electron energy threshold $E_{\mathrm{cut}}$.}
\label{sys4}
\begin{center}
\begin{tabularx}{0.9\linewidth}{l|ZZZZZZZZZ}

\hline \hline
                            & \multicolumn{9}{c}{$M_4$  [$10^{-3}\rm GeV^4$]}\\

$E_{\mathrm{cut}}$[GeV]          &0.4&0.6&0.8&1.0&1.2&1.4&1.6&1.8&2.0\\
\hline\hline
                                                                   Electron Detection&    0.042&    0.119&    0.052&    0.021&    0.005&    0.001&    0.000&    0.000&    0.000\\
 \hline
                                                                $(D^{(*)} e \nu)$ form factors&    0.466&    0.250&    0.186&    0.123&    0.098&    0.052&    0.031&    0.016&    0.007\\
                                                         $\mathcal{B}(D^{(*)} e \nu)$&    0.067&    0.067&    0.015&    0.015&    0.008&    0.004&    0.003&    0.001&    0.001\\
                                                                  $(D^{**}e\nu)$ form factors&    0.519&    0.206&    0.066&    0.017&    0.009&    0.003&    0.001&    0.000&    0.000\\
   $\mathcal{B} (D^{(*)}_{\rm{non-res}}                   \pi e \nu / D^{**}  e \nu)$&    0.483&    0.345&    0.088&    0.026&    0.018&    0.005&    0.003&    0.001&    0.000\\
 \hline
                                                                            Continuum&    0.001&    0.001&    0.000&    0.000&    0.000&    0.000&    0.000&    0.000&    0.000\\
                                                                        $M_{\rm{bc}}$&    0.160&    0.090&    0.021&    0.012&    0.004&    0.002&    0.000&    0.000&    0.000\\
                                                                          $X_u e \nu$&    0.181&    0.138&    0.094&    0.056&    0.030&    0.012&    0.004&    0.001&    0.000\\
                                                                         Hadron Fakes&    0.049&    0.036&    0.019&    0.007&    0.002&    0.000&    0.000&    0.000&    0.000\\
                                                                $B \to D^{(*)} \to e$&    0.079&    0.046&    0.022&    0.008&    0.002&    0.000&    0.000&    0.000&    0.000\\
                                                                          Secondaries&    0.085&    0.050&    0.023&    0.008&    0.002&    0.000&    0.000&    0.000&    0.000\\
 \hline
                                                                            Unfolding&    0.270&    0.157&    0.027&    0.002&    0.014&    0.008&    0.004&    0.000&    0.000\\
 \hline
                                                                    Total Systematics&    0.935&    0.548&    0.247&    0.142&    0.106&    0.055&    0.032&    0.017&    0.007\\
\hline\hline
\end{tabularx}
\end{center}
\end{table}

\begin{table}[htb]
\caption{Breakdown of the systematic errors for the partial branching fractions, $\Delta\mathcal{B}$ for $B \rightarrow X_c e \nu$ in the $B$ meson rest frame for nine values of the electron energy threshold $E_{\mathrm{cut}}$.}
\label{sys0}
\begin{center}
\begin{tabularx}{0.9\linewidth}{l|ZZZZZZZZZ}

\hline \hline
                              & \multicolumn{9}{c}{$\Delta\mathcal{B}$[$10^{-2}$] }  \\
$E_{\mathrm{cut}}$[GeV]          &0.4&0.6&0.8&1.0&1.2&1.4&1.6&1.8&2.0\\
\hline\hline
                                                                     Electron Detection&     0.17&     0.17&     0.15&     0.14&     0.12&     0.10&     0.06&     0.03&     0.01\\
 \hline
                                                                $(D^{(*)} e \nu)$ form factors&     0.01&     0.01&     0.01&     0.01&     0.01&     0.01&     0.01&     0.01&     0.01\\
                                                         $\mathcal{B}(D^{(*)} e \nu)$&     0.05&     0.05&     0.04&     0.04&     0.03&     0.02&     0.02&     0.01&     0.00\\
                                                                  $(D^{**}e\nu)$ form factors&     0.05&     0.04&     0.03&     0.02&     0.01&     0.01&     0.01&     0.00&     0.00\\
   $\mathcal{B} (D^{(*)}_{\rm{non-res}}                   \pi e \nu / D^{**}  e \nu)$&     0.09&     0.09&     0.09&     0.07&     0.05&     0.03&     0.02&     0.01&     0.00\\
 \hline
                                                                            Continuum&     0.04&     0.04&     0.03&     0.03&     0.03&     0.02&     0.01&     0.01&     0.00\\
                                                                        $M_{\rm{bc}}$&     0.04&     0.03&     0.02&     0.01&     0.00&     0.00&     0.00&     0.00&     0.00\\
                                                                          $X_u e \nu$&     0.03&     0.03&     0.03&     0.03&     0.02&     0.02&     0.02&     0.01&     0.01\\
                                                                         Hadron Fakes&     0.02&     0.01&     0.01&     0.01&     0.00&     0.00&     0.00&     0.00&     0.00\\
                                                                $B \to D^{(*)} \to e$&     0.02&     0.02&     0.02&     0.01&     0.01&     0.00&     0.00&     0.00&     0.00\\
                                                                          Secondaries&     0.02&     0.02&     0.01&     0.01&     0.00&     0.00&     0.00&     0.00&     0.00\\
 \hline
                                                                            Unfolding&     0.02&     0.01&     0.00&     0.00&     0.00&     0.00&     0.00&     0.00&     0.00\\
 \hline
                                                                    Total Systematics&     0.22&     0.21&     0.19&     0.17&     0.14&     0.11&     0.07&     0.04&     0.03\\
 
\hline\hline
\end{tabularx}
\end{center}
\end{table}

\clearpage
\begin{sidewaystable}[htb]
\caption{Measured moments, $M_1$, $M_2$, $M_3$, $M_4$ and the partial branching fraction for $B \rightarrow X_c e \nu$ in the $B$ meson rest frame for nine values of the threshold electron energy $E_{\mathrm{cut}}$.  The first error is statistical, and the second error is the systematic.}
\label{moments2}
\begin{center}
\begin{tabular}{r| @{\hspace{0.2cm}} r@{\hspace{0.4cm}} r@{\hspace{0.4cm}} r@{\hspace{0.4cm}} r@{\hspace{0.4cm}} r@{\hspace{0.1cm}}}

\hline \hline
$E_{\mathrm{cut}}$[GeV]& \multicolumn{1}{c}{$M_1$  [MeV]} & \multicolumn{1}{c}{$M_2$  [$10^{-3}\rm GeV^2$]}& \multicolumn{1}{c}{$M_3$  [$10^{-3}\rm GeV^3$]} & \multicolumn{1}{c}{$M_4$  [$10^{-3}\rm GeV^4$]} & \multicolumn{1}{c}{$\Delta\mathcal{B}$  [$10^{-2}$]} \\

\hline\hline
 0.4& 1393.92 $\pm$  6.73 $\pm$  3.02&  168.77 $\pm$  3.68 $\pm$  1.53&  -21.04 $\pm$  1.93 $\pm$  0.66&  64.153 $\pm$ 1.813 $\pm$ 0.935&   10.44 $\pm$  0.19 $\pm$  0.22\\
 0.6& 1427.82 $\pm$  5.82 $\pm$  2.55&  146.15 $\pm$  2.88 $\pm$  1.08&  -11.04 $\pm$  1.35 $\pm$  0.49&  45.366 $\pm$ 1.108 $\pm$ 0.548&   10.07 $\pm$  0.18 $\pm$  0.21\\
 0.8& 1480.04 $\pm$  4.81 $\pm$  2.13&  117.97 $\pm$  2.05 $\pm$  0.55&   -3.45 $\pm$  0.83 $\pm$  0.30&  28.701 $\pm$ 0.585 $\pm$ 0.247&    9.42 $\pm$  0.16 $\pm$  0.19\\
 1.0& 1547.76 $\pm$  3.96 $\pm$  1.45&   88.17 $\pm$  1.42 $\pm$  0.36&    0.83 $\pm$  0.49 $\pm$  0.20&  15.962 $\pm$ 0.302 $\pm$ 0.142&    8.41 $\pm$  0.15 $\pm$  0.17\\
 1.2& 1627.79 $\pm$  3.26 $\pm$  1.08&   61.36 $\pm$  1.02 $\pm$  0.36&    2.40 $\pm$  0.30 $\pm$  0.11&   7.876 $\pm$ 0.162 $\pm$ 0.106&    7.11 $\pm$  0.13 $\pm$  0.14\\
 1.4& 1719.96 $\pm$  2.58 $\pm$  1.10&   38.99 $\pm$  0.71 $\pm$  0.24&    2.33 $\pm$  0.16 $\pm$  0.07&   3.314 $\pm$ 0.080 $\pm$ 0.055&    5.52 $\pm$  0.11 $\pm$  0.11\\
 1.6& 1826.15 $\pm$  1.80 $\pm$  1.03&   21.75 $\pm$  0.47 $\pm$  0.22&    1.45 $\pm$  0.08 $\pm$  0.05&   1.129 $\pm$ 0.033 $\pm$ 0.032&    3.71 $\pm$  0.09 $\pm$  0.07\\
 1.8& 1943.18 $\pm$  0.93 $\pm$  1.16&   10.14 $\pm$  0.28 $\pm$  0.18&    0.68 $\pm$  0.03 $\pm$  0.04&   0.283 $\pm$ 0.010 $\pm$ 0.017&    1.93 $\pm$  0.06 $\pm$  0.04\\
 2.0& 2077.59 $\pm$  0.21 $\pm$  1.23&    3.47 $\pm$  0.13 $\pm$  0.19&    0.19 $\pm$  0.01 $\pm$  0.03&   0.047 $\pm$ 0.002 $\pm$ 0.007&    0.53 $\pm$  0.02 $\pm$  0.02\\
 \hline\hline
\end{tabular}
\end{center}
\end{sidewaystable}

\begin{table}[htb]
\caption{Results and breakdown of the systematic errors for the partial branching fractions of charmed semileptonic $B$ decays, independently measured for $B^+$ and $B^0$ decays with electron energy threshold values of 0.4 GeV and 0.6 GeV.}
\label{bplusbzerosys}
\begin{center}
\begin{tabularx}{0.6\linewidth}{l|ZZZZ}
\hline \hline
                              & \multicolumn{2}{c}{$\Delta\mathcal{B}(B^+)$[$10^{-2}$] } & \multicolumn{2}{c}{$\Delta\mathcal{B}(B^0)$[$10^{-2}$] }  \\
$E_{\mathrm{cut}}$[GeV]                             					 &0.4    &    0.6    &    0.4    &   0.6\\
\hline\hline
Electron Detection  									 & 0.18& 0.17 & 0.17& 0.16\\    
\hline
$(D^{(*)} e \nu)$ form factors							&   0.01& 0.01& 0.01& 0.01\\
$\mathcal{B}(D^{(*)} e \nu)$							 &   0.08& 0.07& 0.02& 0.02\\
$(D^{**}e\nu)$ form factors							&   0.05& 0.04& 0.04& 0.03\\  
$\mathcal{B} (D^{(*)}_{\rm{non-res}} \pi e \nu / D^{**}  e \nu)$ 	 &   0.09& 0.09& 0.10& 0.10\\  
\hline
Continuum										 &   0.10& 0.09& 0.04& 0.04\\
$M_{\rm{bc}}$										 &   0.10& 0.08& 0.04& 0.03\\
$X_u e \nu$ 										 &   0.04& 0.04& 0.03& 0.03\\
Hadron fakes 										 &   0.02& 0.02& 0.02& 0.02\\
$B \to D^{(*)} \to e$									 &   0.00& 0.00& 0.04& 0.04\\    
Secondaries 										 &   0.08& 0.06& 0.02& 0.02\\
\hline						
Unfolding											 &   0.02 & 0.01& 0.06& 0.06\\
\hline			
$\Delta\mathcal{B}$									&   10.79  & 10.34 &10.08 &9.80  \\
$\pm$ (stat.)										 &   0.25 & 0.23      & 0.30&0.29\\
$\pm$ (sys.)										 &   0.27 & 0.25      & 0.22& 0.21\\
\hline\hline
\end{tabularx}
\end{center}
\end{table}

\begin{table}[htb]
\caption{Correlation coefficients between $M_1$ measurements, $\rho_{M_1,M_1}$.}
\begin{center}
\begin{tabularx}{0.75\linewidth}{X|ZZZZZZZZZ}
\hline\hline
          &$M_1^{0.4}$&$M_1^{0.6}$&$M_1^{0.8}$&$M_1^{1.0}$&$M_1^{1.2}$&$M_1^{1.4}$&$M_1^{1.6}$&$M_1^{1.8}$&$M_1^{2.0}$\\ 
\hline
$M_1^{0.4}$&  1.00&  0.97&  0.86&  0.65&  0.47&  0.33&  0.25&  0.10&  0.05\\
$M_1^{0.6}$&  0.97&  1.00&  0.94&  0.76&  0.59&  0.42&  0.31&  0.14&  0.13\\
$M_1^{0.8}$&  0.86&  0.94&  1.00&  0.92&  0.77&  0.56&  0.41&  0.16&  0.15\\
$M_1^{1.0}$&  0.65&  0.76&  0.92&  1.00&  0.92&  0.73&  0.55&  0.26&  0.11\\
$M_1^{1.2}$&  0.47&  0.59&  0.77&  0.92&  1.00&  0.90&  0.72&  0.40&  0.19\\
$M_1^{1.4}$&  0.33&  0.42&  0.56&  0.73&  0.90&  1.00&  0.93&  0.67&  0.36\\
$M_1^{1.6}$&  0.25&  0.31&  0.41&  0.55&  0.72&  0.93&  1.00&  0.83&  0.46\\
$M_1^{1.8}$&  0.10&  0.14&  0.16&  0.26&  0.40&  0.67&  0.83&  1.00&  0.68\\
$M_1^{2.0}$&  0.05&  0.13&  0.15&  0.11&  0.19&  0.36&  0.46&  0.68&  1.00\\

\hline
\end{tabularx}
\end{center}
\label{m1m1}
\end{table}%

\begin{table}[htb]
\caption{Correlation coefficients between $M_1$ and $M_2$ measurements, $\rho_{M_1,M_2}$.}
\begin{center}
\begin{tabularx}{0.75\linewidth}{X|ZZZZZZZZZ}

\hline\hline
          &$M_2^{0.4}$&$M_2^{0.6}$&$M_2^{0.8}$&$M_2^{1.0}$&$M_2^{1.2}$&$M_2^{1.4}$&$M_2^{1.6}$&$M_2^{1.8}$&$M_2^{2.0}$\\ 
\hline
$M_1^{0.4}$& $-$0.51& $-$0.42& $-$0.27&  0.75& $-$0.23&  0.11&  0.17&  0.18&  0.09\\
$M_1^{0.6}$& $-$0.44& $-$0.37& $-$0.26&  0.67& $-$0.11&  0.17&  0.21&  0.20&  0.11\\
$M_1^{0.8}$& $-$0.26& $-$0.22& $-$0.18&  0.72&  0.08&  0.24&  0.27&  0.23&  0.12\\
$M_1^{1.0}$&  0.01&  0.04&  0.03&  0.11&  0.26&  0.39&  0.42&  0.34&  0.05\\
$M_1^{1.2}$&  0.31&  0.33&  0.30&  0.27&  0.32&  0.47&  0.50&  0.46&  0.20\\
$M_1^{1.4}$&  0.47&  0.51&  0.53&  0.53&  0.51&  0.56&  0.60&  0.53&  0.25\\
$M_1^{1.6}$&  0.50&  0.56&  0.59&  0.64&  0.66&  0.69&  0.68&  0.66&  0.40\\
$M_1^{1.8}$&  0.43&  0.49&  0.57&  0.65&  0.71&  0.80&  0.80&  0.75&  0.56\\
$M_1^{2.0}$&  0.28&  0.34&  0.37&  0.43&  0.48&  0.64&  0.68&  0.80&  0.84\\

\hline
\end{tabularx}
\end{center}
\label{m1m2}
\end{table}%

\begin{table}[htb]
\caption{Correlation coefficients between $M_1$ and $M_3$ measurements, $\rho_{M_1,M_3}$.}
\begin{center}
\begin{tabularx}{0.75\linewidth}{X|ZZZZZZZZZ}
\hline\hline
          &$M_3^{0.4}$&$M_3^{0.6}$&$M_3^{0.8}$&$M_3^{1.0}$&$M_3^{1.2}$&$M_3^{1.4}$&$M_3^{1.6}$&$M_3^{1.8}$&$M_3^{2.0}$\\ 
\hline
$M_1^{0.4}$&  0.77&  0.71&  0.57&  0.35&  0.24&  0.27&  0.24&  0.30&  0.10\\
$M_1^{0.6}$&  0.76&  0.74&  0.64&  0.47&  0.34&  0.34&  0.28&  0.33&  0.10\\
$M_1^{0.8}$&  0.72&  0.73&  0.72&  0.66&  0.54&  0.47&  0.40&  0.44&  0.18\\
$M_1^{1.0}$&  0.55&  0.59&  0.70&  0.77&  0.74&  0.65&  0.58&  0.40&  0.09\\
$M_1^{1.2}$&  0.36&  0.44&  0.60&  0.76&  0.84&  0.83&  0.73&  0.53&  0.30\\
$M_1^{1.4}$&  0.29&  0.36&  0.48&  0.63&  0.78&  0.87&  0.83&  0.71&  0.45\\
$M_1^{1.6}$&  0.23&  0.30&  0.41&  0.51&  0.62&  0.80&  0.84&  0.72&  0.63\\
$M_1^{1.8}$&  0.17&  0.26&  0.37&  0.42&  0.48&  0.66&  0.73&  0.69&  0.80\\
$M_1^{2.0}$&  0.04&  0.11&  0.20&  0.21&  0.23&  0.39&  0.45&  0.57&  0.92\\

\hline
\end{tabularx}
\end{center}
\label{m1m3}
\end{table}%

\begin{table}[htb]
\caption{Correlation coefficients between $M_1$ and $M_4$ measurements, $\rho_{M_1,M_4}$.}
\begin{center}
\begin{tabularx}{0.75\linewidth}{X|ZZZZZZZZZ}
\hline\hline
          &$M_4^{0.4}$&$M_4^{0.6}$&$M_4^{0.8}$&$M_4^{1.0}$&$M_4^{1.2}$&$M_4^{1.4}$&$M_4^{1.6}$&$M_4^{1.8}$&$M_4^{2.0}$\\ 
\hline
$M_1^{0.4}$& $-$0.58& $-$0.49& $-$0.34&  0.42& $-$0.53&  0.05&  0.17&  0.24&  0.10\\
$M_1^{0.6}$& $-$0.50& $-$0.43& $-$0.30&  0.61& $-$0.13&  0.19&  0.26&  0.32&  0.20\\
$M_1^{0.8}$& $-$0.29& $-$0.23& $-$0.18& $-$0.15&  0.19&  0.34&  0.37&  0.44&  0.25\\
$M_1^{1.0}$&  0.06&  0.07&  0.05&  0.19&  0.38&  0.53&  0.59&  0.37&  0.07\\
$M_1^{1.2}$&  0.30&  0.34&  0.33&  0.34&  0.43&  0.63&  0.68&  0.52&  0.44\\
$M_1^{1.4}$&  0.41&  0.45&  0.50&  0.52&  0.56&  0.70&  0.76&  0.62&  0.55\\
$M_1^{1.6}$&  0.37&  0.44&  0.46&  0.53&  0.57&  0.70&  0.73&  0.74&  0.72\\
$M_1^{1.8}$&  0.26&  0.31&  0.39&  0.43&  0.47&  0.62&  0.66&  0.70&  0.85\\
$M_1^{2.0}$&  0.16&  0.21&  0.23&  0.25&  0.27&  0.43&  0.48&  0.60&  0.93\\

 \hline
\end{tabularx}
\end{center}
\label{m1m4}
\end{table}%

\begin{table}[htb]
\caption{Correlation coefficients between $M_1$ and $\Delta\mathcal{B}$ measurements, $\rho_{M_1,\Delta\mathcal{B}}$.}
\begin{center}
\begin{tabularx}{0.75\linewidth}{X|ZZZZZZZZZ}
\hline\hline
          &$\Delta\mathcal{B}^{0.4}$&$\Delta\mathcal{B}^{0.6}$&$\Delta\mathcal{B}^{0.8}$&$\Delta\mathcal{B}^{1.0}$&$\Delta\mathcal{B}^{1.2}$&$\Delta\mathcal{B}^{1.4}$&$\Delta\mathcal{B}^{1.6}$&$\Delta\mathcal{B}^{1.8}$&$\Delta\mathcal{B}^{2.0}$\\ 
\hline
$M_1^{0.4}$& $-$0.36& $-$0.36& $-$0.31& $-$0.17& $-$0.08& $-$0.04& $-$0.03& $-$0.08& $-$0.04\\
$M_1^{0.6}$& $-$0.30& $-$0.31& $-$0.28& $-$0.17& $-$0.10& $-$0.05& $-$0.02& $-$0.10& $-$0.05\\
$M_1^{0.8}$& $-$0.14& $-$0.18& $-$0.21& $-$0.16& $-$0.11& $-$0.06& $-$0.02& $-$0.08& $-$0.02\\
$M_1^{1.0}$& $-$0.09& $-$0.12& $-$0.13& $-$0.07& $-$0.08& $-$0.06& $-$0.04& $-$0.04& $-$0.02\\
$M_1^{1.2}$&  0.10&  0.08&  0.07&  0.07&  0.01& $-$0.07& $-$0.18& $-$0.02&  0.00\\
$M_1^{1.4}$&  0.26&  0.26&  0.27&  0.28&  0.22&  0.09&  0.00& $-$0.09&  0.03\\
$M_1^{1.6}$&  0.44&  0.47&  0.47&  0.50&  0.47&  0.38&  0.24&  0.09&  0.07\\
$M_1^{1.8}$&  0.42&  0.47&  0.55&  0.64&  0.69&  0.63&  0.57&  0.41&  0.17\\
$M_1^{2.0}$&  0.32&  0.36&  0.43&  0.55&  0.63&  0.62&  0.63&  0.59&  0.31\\

\hline
\end{tabularx}
\end{center}
\label{m1m0}
\end{table}%

\begin{table}[htb]
\caption{Correlation coefficients between $M_2$ measurements, $\rho_{M_2,M_2}$.}
\begin{center}
\begin{tabularx}{0.75\linewidth}{X|ZZZZZZZZZ}
\hline\hline
          &$M_2^{0.4}$&$M_2^{0.6}$&$M_2^{0.8}$&$M_2^{1.0}$&$M_2^{1.2}$&$M_2^{1.4}$&$M_2^{1.6}$&$M_2^{1.8}$&$M_2^{2.0}$\\ 
\hline
$M_2^{0.4}$&  1.00&  0.95&  0.83&  0.58&  0.41&  0.38&  0.37&  0.35&  0.26\\
$M_2^{0.6}$&  0.95&  1.00&  0.91&  0.69&  0.52&  0.44&  0.41&  0.40&  0.29\\
$M_2^{0.8}$&  0.83&  0.91&  1.00&  0.90&  0.73&  0.61&  0.57&  0.52&  0.35\\
$M_2^{1.0}$&  0.58&  0.69&  0.90&  1.00&  0.92&  0.79&  0.73&  0.68&  0.48\\
$M_2^{1.2}$&  0.41&  0.52&  0.73&  0.92&  1.00&  0.94&  0.85&  0.80&  0.58\\
$M_2^{1.4}$&  0.38&  0.44&  0.61&  0.79&  0.94&  1.00&  0.95&  0.87&  0.63\\
$M_2^{1.6}$&  0.37&  0.41&  0.57&  0.73&  0.85&  0.95&  1.00&  0.95&  0.74\\
$M_2^{1.8}$&  0.35&  0.40&  0.52&  0.68&  0.80&  0.87&  0.95&  1.00&  0.88\\
$M_2^{2.0}$&  0.26&  0.29&  0.35&  0.48&  0.58&  0.63&  0.74&  0.88&  1.00\\
 
\hline
\end{tabularx}
\end{center}
\label{m2m2}
\end{table}%

\begin{table}[htb]
\caption{Correlation coefficients between $M_2$ and $M_3$ measurements, $\rho_{M_2,M_3}$.}
\begin{center}
\begin{tabularx}{0.75\linewidth}{X|ZZZZZZZZZ}
\hline\hline
          &$M_3^{0.4}$&$M_3^{0.6}$&$M_3^{0.8}$&$M_3^{1.0}$&$M_3^{1.2}$&$M_3^{1.4}$&$M_3^{1.6}$&$M_3^{1.8}$&$M_3^{2.0}$\\ 
\hline
$M_2^{0.4}$& $-$0.55& $-$0.49& $-$0.29& $-$0.51&  0.26&  0.39&  0.40&  0.37&  0.28\\
$M_2^{0.6}$& $-$0.52& $-$0.47& $-$0.26& $-$0.09&  0.30&  0.44&  0.43&  0.45&  0.32\\
$M_2^{0.8}$& $-$0.29& $-$0.25& $-$0.13&  0.11&  0.40&  0.55&  0.60&  0.58&  0.43\\
$M_2^{1.0}$&  0.49&  0.64& $-$0.07&  0.22&  0.44&  0.62&  0.71&  0.71&  0.55\\
$M_2^{1.2}$&  0.03&  0.13&  0.27&  0.38&  0.49&  0.65&  0.81&  0.80&  0.65\\
$M_2^{1.4}$&  0.18&  0.28&  0.39&  0.51&  0.59&  0.67&  0.81&  0.86&  0.76\\
$M_2^{1.6}$&  0.17&  0.28&  0.41&  0.54&  0.64&  0.70&  0.81&  0.89&  0.89\\
$M_2^{1.8}$&  0.09&  0.18&  0.33&  0.46&  0.57&  0.63&  0.74&  0.84&  0.96\\
$M_2^{2.0}$& $-$0.03&  0.05&  0.16&  0.24&  0.32&  0.34&  0.47&  0.67&  0.92\\

\hline
\end{tabularx}
\end{center}
\label{m2m3}
\end{table}%

  \begin{table}[htb]
\caption{Correlation coefficients between $M_2$ and $M_4$ measurements, $\rho_{M_2,M_4}$.}
\begin{center}
\begin{tabularx}{0.75\linewidth}{X|ZZZZZZZZZ}
\hline\hline
          &$M_4^{0.4}$&$M_4^{0.6}$&$M_4^{0.8}$&$M_4^{1.0}$&$M_4^{1.2}$&$M_4^{1.4}$&$M_4^{1.6}$&$M_4^{1.8}$&$M_4^{2.0}$\\ 
\hline
$M_2^{0.4}$&  0.98&  0.93&  0.74&  0.45&  0.29&  0.26&  0.26&  0.27&  0.28\\
$M_2^{0.6}$&  0.94&  0.95&  0.83&  0.58&  0.42&  0.36&  0.33&  0.36&  0.32\\
$M_2^{0.8}$&  0.84&  0.89&  0.97&  0.84&  0.69&  0.59&  0.59&  0.55&  0.45\\
$M_2^{1.0}$&  0.59&  0.67&  0.86&  0.95&  0.90&  0.81&  0.80&  0.78&  0.67\\
$M_2^{1.2}$&  0.41&  0.49&  0.67&  0.83&  0.93&  0.90&  0.90&  0.87&  0.78\\
$M_2^{1.4}$&  0.37&  0.42&  0.55&  0.72&  0.86&  0.91&  0.95&  0.93&  0.85\\
$M_2^{1.6}$&  0.32&  0.35&  0.46&  0.61&  0.73&  0.80&  0.90&  0.94&  0.94\\
$M_2^{1.8}$&  0.26&  0.28&  0.34&  0.47&  0.58&  0.63&  0.74&  0.86&  0.98\\
$M_2^{2.0}$&  0.17&  0.16&  0.19&  0.28&  0.36&  0.39&  0.52&  0.71&  0.94\\

\hline

\end{tabularx}
\end{center}
\label{m2m4}
\end{table}%

\begin{table}[htb]
\caption{Correlation coefficients between $M_2$ and $\Delta\mathcal{B}$ measurements, $\rho_{M_2,\Delta\mathcal{B}}$.}
\begin{center}
\begin{tabularx}{0.75\linewidth}{X|ZZZZZZZZZ}
\hline\hline
          &$\Delta\mathcal{B}^{0.4}$&$\Delta\mathcal{B}^{0.6}$&$\Delta\mathcal{B}^{0.8}$&$\Delta\mathcal{B}^{1.0}$&$\Delta\mathcal{B}^{1.2}$&$\Delta\mathcal{B}^{1.4}$&$\Delta\mathcal{B}^{1.6}$&$\Delta\mathcal{B}^{1.8}$&$\Delta\mathcal{B}^{2.0}$\\ 
\hline
$M_2^{0.4}$&  0.54&  0.57&  0.58&  0.53&  0.41&  0.35&  0.26&  0.20&  0.12\\
$M_2^{0.6}$&  0.47&  0.52&  0.56&  0.54&  0.44&  0.37&  0.26&  0.20&  0.12\\
$M_2^{0.8}$&  0.49&  0.54&  0.59&  0.59&  0.52&  0.41&  0.35&  0.24&  0.16\\
$M_2^{1.0}$&  0.29&  0.33&  0.43&  0.51&  0.53&  0.47&  0.40&  0.26&  0.17\\
$M_2^{1.2}$&  0.29&  0.35&  0.42&  0.51&  0.56&  0.52&  0.47&  0.32&  0.19\\
$M_2^{1.4}$&  0.31&  0.37&  0.41&  0.48&  0.55&  0.58&  0.57&  0.41&  0.26\\
$M_2^{1.6}$&  0.27&  0.31&  0.39&  0.49&  0.56&  0.62&  0.66&  0.54&  0.34\\
$M_2^{1.8}$&  0.23&  0.29&  0.42&  0.56&  0.63&  0.69&  0.74&  0.70&  0.48\\
$M_2^{2.0}$&  0.25&  0.33&  0.42&  0.56&  0.61&  0.72&  0.79&  0.71&  0.60\\

\hline

\end{tabularx}
\end{center}
\label{m2m0}
\end{table}%

\begin{table}[htb]
\caption{Correlation coefficients between $M_3$ measurements, $\rho_{M_3,M_3}$.}
\begin{center}
\begin{tabularx}{0.75\linewidth}{X|ZZZZZZZZZ}
\hline\hline
          &$M_3^{0.4}$&$M_3^{0.6}$&$M_3^{0.8}$&$M_3^{1.0}$&$M_3^{1.2}$&$M_3^{1.4}$&$M_3^{1.6}$&$M_3^{1.8}$&$M_3^{2.0}$\\ 
\hline
$M_3^{0.4}$&  1.00&  0.97&  0.86&  0.57&  0.39&  0.35&  0.29&  0.14&  0.02\\
$M_3^{0.6}$&  0.97&  1.00&  0.94&  0.71&  0.52&  0.46&  0.41&  0.23&  0.10\\
$M_3^{0.8}$&  0.86&  0.94&  1.00&  0.90&  0.73&  0.63&  0.57&  0.39&  0.21\\
$M_3^{1.0}$&  0.57&  0.71&  0.90&  1.00&  0.93&  0.81&  0.72&  0.51&  0.30\\
$M_3^{1.2}$&  0.39&  0.52&  0.73&  0.93&  1.00&  0.94&  0.85&  0.63&  0.38\\
$M_3^{1.4}$&  0.35&  0.46&  0.63&  0.81&  0.94&  1.00&  0.95&  0.75&  0.49\\
$M_3^{1.6}$&  0.29&  0.41&  0.57&  0.72&  0.85&  0.95&  1.00&  0.89&  0.67\\
$M_3^{1.8}$&  0.14&  0.23&  0.39&  0.51&  0.63&  0.75&  0.89&  1.00&  0.89\\
$M_3^{2.0}$&  0.02&  0.10&  0.21&  0.30&  0.38&  0.49&  0.67&  0.89&  1.00\\

\hline

\end{tabularx}
\end{center}
\label{m3m3}
\end{table}%

\begin{table}[htb]
\caption{Correlation coefficients between $M_3$ and $M_4$ measurements, $\rho_{M_3,M_4}$.}
\begin{center}
\begin{tabularx}{0.75\linewidth}{X|ZZZZZZZZZ}
\hline\hline
          &$M_4^{0.4}$&$M_4^{0.6}$&$M_4^{0.8}$&$M_4^{1.0}$&$M_4^{1.2}$&$M_4^{1.4}$&$M_4^{1.6}$&$M_4^{1.8}$&$M_4^{2.0}$\\ 
\hline
$M_3^{0.4}$& $-$0.65& $-$0.57& $-$0.35&  0.35& $-$0.18&  0.08&  0.07& $-$0.04& $-$0.13\\
$M_3^{0.6}$& $-$0.55& $-$0.49& $-$0.28&  0.70&  0.08&  0.26&  0.26&  0.16&  0.05\\
$M_3^{0.8}$& $-$0.30& $-$0.27& $-$0.09&  0.14&  0.37&  0.49&  0.53&  0.42&  0.27\\
$M_3^{1.0}$&  0.04&  0.11&  0.21&  0.35&  0.54&  0.68&  0.73&  0.63&  0.43\\
$M_3^{1.2}$&  0.31&  0.35&  0.46&  0.56&  0.65&  0.77&  0.85&  0.76&  0.55\\
$M_3^{1.4}$&  0.38&  0.44&  0.54&  0.69&  0.76&  0.83&  0.90&  0.84&  0.61\\
$M_3^{1.6}$&  0.34&  0.39&  0.52&  0.68&  0.77&  0.84&  0.94&  0.93&  0.78\\
$M_3^{1.8}$&  0.27&  0.30&  0.40&  0.51&  0.62&  0.72&  0.87&  0.98&  0.93\\
$M_3^{2.0}$&  0.16&  0.17&  0.25&  0.34&  0.42&  0.53&  0.71&  0.92&  0.99\\

\hline

\end{tabularx}
\end{center}
\label{m3m4}
\end{table}%

\begin{table}[htb]
\caption{Correlation coefficients between $M_3$ and $\Delta\mathcal{B}$  measurements, $\rho_{M_3,\Delta\mathcal{B}}$.}
\begin{center}
\begin{tabularx}{0.75\linewidth}{X|ZZZZZZZZZ}
\hline\hline
          &$\Delta\mathcal{B}^{0.4}$&$\Delta\mathcal{B}^{0.6}$&$\Delta\mathcal{B}^{0.8}$&$\Delta\mathcal{B}^{1.0}$&$\Delta\mathcal{B}^{1.2}$&$\Delta\mathcal{B}^{1.4}$&$\Delta\mathcal{B}^{1.6}$&$\Delta\mathcal{B}^{1.8}$&$\Delta\mathcal{B}^{2.0}$\\ 
\hline
$M_3^{0.4}$& $-$0.14& $-$0.13& $-$0.10&  0.00&  0.09& $-$0.08&  0.03&  0.06&  0.00\\
$M_3^{0.6}$& $-$0.06& $-$0.06& $-$0.05&  0.01&  0.08&  0.01&  0.05&  0.08&  0.01\\
$M_3^{0.8}$&  0.08&  0.06&  0.02&  0.02&  0.06&  0.05&  0.08&  0.10&  0.05\\
$M_3^{1.0}$& $-$0.02&  0.02&  0.06&  0.09&  0.06&  0.07&  0.10&  0.10&  0.06\\
$M_3^{1.2}$&  0.25&  0.25&  0.25&  0.21&  0.14&  0.10&  0.13&  0.13&  0.09\\
$M_3^{1.4}$&  0.23&  0.29&  0.32&  0.34&  0.28&  0.18&  0.17&  0.14&  0.11\\
$M_3^{1.6}$&  0.28&  0.35&  0.42&  0.49&  0.47&  0.38&  0.32&  0.23&  0.14\\
$M_3^{1.8}$&  0.32&  0.37&  0.47&  0.55&  0.61&  0.59&  0.57&  0.38&  0.21\\
$M_3^{2.0}$&  0.30&  0.36&  0.46&  0.56&  0.65&  0.66&  0.72&  0.46&  0.31\\

\hline

\end{tabularx}
\end{center}
\label{m3m0}
\end{table}%

\begin{table}[htb]
\caption{Correlation coefficients between $M_4$ measurements, $\rho_{M_4,M_4}$.}
\begin{center}
\begin{tabularx}{0.75\linewidth}{X|ZZZZZZZZZ}
\hline\hline
          &$M_4^{0.4}$&$M_4^{0.6}$&$M_4^{0.8}$&$M_4^{1.0}$&$M_4^{1.2}$&$M_4^{1.4}$&$M_4^{1.6}$&$M_4^{1.8}$&$M_4^{2.0}$\\ 
\hline
$M_4^{0.4}$&  1.00&  0.97&  0.78&  0.51&  0.36&  0.31&  0.29&  0.28&  0.23\\
$M_4^{0.6}$&  0.97&  1.00&  0.86&  0.62&  0.46&  0.39&  0.36&  0.31&  0.23\\
$M_4^{0.8}$&  0.78&  0.86&  1.00&  0.87&  0.70&  0.61&  0.55&  0.43&  0.30\\
$M_4^{1.0}$&  0.51&  0.62&  0.87&  1.00&  0.92&  0.84&  0.78&  0.65&  0.47\\
$M_4^{1.2}$&  0.36&  0.46&  0.70&  0.92&  1.00&  0.96&  0.90&  0.77&  0.59\\
$M_4^{1.4}$&  0.31&  0.39&  0.61&  0.84&  0.96&  1.00&  0.96&  0.84&  0.65\\
$M_4^{1.6}$&  0.29&  0.36&  0.55&  0.78&  0.90&  0.96&  1.00&  0.94&  0.81\\
$M_4^{1.8}$&  0.28&  0.31&  0.43&  0.65&  0.77&  0.84&  0.94&  1.00&  0.96\\
$M_4^{2.0}$&  0.23&  0.23&  0.30&  0.47&  0.59&  0.65&  0.81&  0.96&  1.00\\

\hline

\end{tabularx}
\end{center}
\label{m4m4}
\end{table}%

\begin{table}[htb]
\caption{Correlation coefficients between $M_4$ and $\Delta\mathcal{B}$ measurements, $\rho_{M_4,\Delta\mathcal{B}}$.}
\begin{center}
\begin{tabularx}{0.75\linewidth}{X|ZZZZZZZZZ}
\hline\hline
          &$\Delta\mathcal{B}^{0.4}$&$\Delta\mathcal{B}^{0.6}$&$\Delta\mathcal{B}^{0.8}$&$\Delta\mathcal{B}^{1.0}$&$\Delta\mathcal{B}^{1.2}$&$\Delta\mathcal{B}^{1.4}$&$\Delta\mathcal{B}^{1.6}$&$\Delta\mathcal{B}^{1.8}$&$\Delta\mathcal{B}^{2.0}$\\ 
\hline
$M_4^{0.4}$&  0.44&  0.50&  0.51&  0.44&  0.29&  0.22&  0.16&  0.09&  0.07\\
$M_4^{0.6}$&  0.37&  0.44&  0.49&  0.45&  0.31&  0.23&  0.16&  0.09&  0.06\\
$M_4^{0.8}$&  0.38&  0.46&  0.52&  0.49&  0.39&  0.26&  0.22&  0.14&  0.08\\
$M_4^{1.0}$&  0.22&  0.24&  0.33&  0.40&  0.39&  0.30&  0.25&  0.14&  0.09\\
$M_4^{1.2}$&  0.15&  0.16&  0.25&  0.35&  0.40&  0.34&  0.29&  0.17&  0.11\\
$M_4^{1.4}$&  0.16&  0.26&  0.32&  0.40&  0.42&  0.38&  0.36&  0.21&  0.14\\
$M_4^{1.6}$&  0.15&  0.20&  0.32&  0.43&  0.46&  0.45&  0.45&  0.29&  0.18\\
$M_4^{1.8}$&  0.11&  0.20&  0.37&  0.55&  0.58&  0.60&  0.60&  0.40&  0.25\\
$M_4^{2.0}$&  0.12&  0.21&  0.37&  0.57&  0.61&  0.66&  0.70&  0.44&  0.34\\

\hline

\end{tabularx}
\end{center}
\label{m4m0}
\end{table}%

 \begin{table}[htb]
\caption{Correlation coefficients between $\Delta\mathcal{B}$ measurements, $\rho_{\Delta\mathcal{B},\Delta\mathcal{B}}$.}
\begin{center}
\begin{tabularx}{0.75\linewidth}{X|ZZZZZZZZZ}
\hline\hline
          &$\Delta\mathcal{B}^{0.4}$&$\Delta\mathcal{B}^{0.6}$&$\Delta\mathcal{B}^{0.8}$&$\Delta\mathcal{B}^{1.0}$&$\Delta\mathcal{B}^{1.2}$&$\Delta\mathcal{B}^{1.4}$&$\Delta\mathcal{B}^{1.6}$&$\Delta\mathcal{B}^{1.8}$&$\Delta\mathcal{B}^{2.0}$\\ 
\hline
$\Delta\mathcal{B}^{0.4}$&  1.00&  0.99&  0.97&  0.93&  0.87&  0.80&  0.71&  0.58&  0.46\\
$\Delta\mathcal{B}^{0.6}$&  0.99&  1.00&  0.99&  0.95&  0.90&  0.83&  0.74&  0.61&  0.48\\
$\Delta\mathcal{B}^{0.8}$&  0.97&  0.99&  1.00&  0.98&  0.94&  0.86&  0.77&  0.64&  0.50\\
$\Delta\mathcal{B}^{1.0}$&  0.93&  0.95&  0.98&  1.00&  0.98&  0.92&  0.83&  0.69&  0.53\\
$\Delta\mathcal{B}^{1.2}$&  0.87&  0.90&  0.94&  0.98&  1.00&  0.97&  0.89&  0.74&  0.57\\
$\Delta\mathcal{B}^{1.4}$&  0.80&  0.83&  0.86&  0.92&  0.97&  1.00&  0.96&  0.82&  0.62\\
$\Delta\mathcal{B}^{1.6}$&  0.71&  0.74&  0.77&  0.83&  0.89&  0.96&  1.00&  0.93&  0.74\\
$\Delta\mathcal{B}^{1.8}$&  0.58&  0.61&  0.64&  0.69&  0.74&  0.82&  0.93&  1.00&  0.90\\
$\Delta\mathcal{B}^{2.0}$&  0.46&  0.48&  0.50&  0.53&  0.57&  0.62&  0.74&  0.90&  1.00\\

\hline
\end{tabularx}
\end{center}
\label{m0m0}
\end{table}%

\clearpage

\section{Acknowledgments}
We thank the KEKB group for the excellent operation of the
accelerator, the KEK cryogenics group for the efficient
operation of the solenoid, and the KEK computer group and
the National Institute of Informatics for valuable computing
and Super-SINET network support. We acknowledge support from
the Ministry of Education, Culture, Sports, Science, and
Technology of Japan and the Japan Society for the Promotion
of Science; the Australian Research Council and the
Australian Department of Education, Science and Training;
the National Science Foundation of China and the Knowledge
Innovation Program of the Chinese Academy of Sciencies under
contract No.~10575109 and IHEP-U-503; the Department of
Science and Technology of India; 
the BK21 program of the Ministry of Education of Korea, 
the CHEP SRC program and Basic Research program 
(grant No.~R01-2005-000-10089-0) of the Korea Science and
Engineering Foundation, and the Pure Basic Research Group 
program of the Korea Research Foundation; 
the Polish State Committee for Scientific Research; 
the Ministry of Science and Technology of the Russian
Federation; the Slovenian Research Agency;  the Swiss
National Science Foundation; the National Science Council
and the Ministry of Education of Taiwan; and the U.S.\
Department of Energy.


\begin{thebibliography}{99}
\bibitem{cc} Throughout this paper, the inclusion of the charge conjugate mode decay is implied unless otherwise stated.
\end{thebibliography}
\end{document}